\documentstyle[aps,eqsecnum,prd,amsfonts,preprint,tighten]{revtex}
\begin{document}
\draft
\title{Observing binary inspiral in\protect\\
gravitational radiation: One interferometer}
\author{Lee Samuel Finn}
\address{Department of Physics and Astronomy,\\
Northwestern University,
Evanston, Illinois\ \ 60208}
\author{David F.\ Chernoff}
\address{Department of Astronomy, Cornell University,\\
Ithaca, New York\ \ 14853}
\date{\today}
\maketitle
\begin{abstract}
Close binary systems of compact objects with less than ten
minutes remaining before coalescence are readily identifiable
sources of gravitational radiation for the United States Laser
Interferometer Gravitational-wave Observatory (LIGO) and the
French-Italian VIRGO gravitational-wave observatory. As a start
toward assessing the full capabilities of the LIGO/VIRGO detector
network, we investigate the sensitivity of individual
LIGO/VIRGO-like interferometers and the precision with which they
can determine the characteristics of an inspiralling binary
system. Since the two interferometers of the LIGO detector share
nearly the same orientation, their joint sensitivity is similar
to that of a single, more sensitive interferometer. We express
our results for a single interferometer of both initial and
advanced LIGO design, and also for the LIGO {\em detector\/} in
the limit that its two interferometers share exactly the same
orientation.

We approximate the secular evolution of a binary system as driven
exclusively by its leading order quadrupole gravitational
radiation. Observations of a binary in a single interferometer
are described by four characteristic quantities: an amplitude
${\cal A}$, a chirp mass ${\cal M}$, a time $T$, and a phase
$\psi$. We find the amplitude signal-to-noise ratio (SNR) $\rho$
of an observed binary system as a function of ${\cal A}$ and
${\cal M}$ for a particular orientation of the binary with
respect to the interferometer, and also the distribution of SNRs
for randomly oriented binaries at a constant distance To assess
the interferometer sensitivity, we calculate the rate at which
sources are expected to be observed and the range to which they
are observable.  Assuming a conservative rate density for
coalescing neutron star binary systems of $8\times10^{-8}\,{\rm
yr}^{-1}\,{\rm Mpc}^{-3}$, we find that the advanced LIGO
detector will observe approximately 69~yr${}^{-1}$ with an
amplitude SNR greater than 8.  Of these, approximately
7~yr${}^{-1}$ will be from binaries at distances greater than
950~Mpc.

We give analytic and numerical results for the precision with
which each of the characteristic quantities can be determined by
interferometer observations. For neutron star binaries, the
fractional one-sigma statistical error in the determination of
${\cal A}$ is equal to $1/\rho$.  For $\rho>8$, the fractional
one-sigma error in the measurement of ${\cal M}$ in the advanced
LIGO detectors is less than $2\times10^{-5}$, a phenomenal
precision. The characteristic time is related to the moment when
coalescence occurs, and can be measured in the advanced detectors
with a one-sigma uncertainty of less than $3\times10^{-4}$~s
(assuming $\rho>8$).

We also explore the sensitivity of these results to a tunable
parameter in the interferometer design (the recycling frequency).
The optimum choice of the parameter is dependent on the goal of
the observations, {\em e.g.,\/} maximizing the rate of detections
or maximizing the precision of measurement. We determine the
optimum parameter values for these two cases.

The calculations leading to the SNR and the precision of
measurement assume that the interferometer observations extend
over only the last several minutes of binary inspiral, during
which time the orbital frequency increases from approximately
5~Hz to 500~Hz. We examine the sensitivity of our results to the
elapsed time of the observation and show that observations of
longer duration lead to very little improvement in the SNR or the
precision of measurement.

\end{abstract}

\narrowtext
\section{Introduction and Motivation}
\label{sec:intro}

Both the United States Laser Interferometer Gravitational-wave
Observatory (LIGO\cite{vogt91,abramovici92}) and the
French/Italian VIRGO gravitational-wave
observatory\cite{bradaschia90} are expected to begin operation in
the late 1990s. Inspiralling binary systems of compact objects
--- either neutron stars or stellar mass black holes with orbital
frequencies ranging from 5~s${}^{-1}$ to 500~s${}^{-1}$ --- are
currently regarded as the most certain observable source for
these detectors: the density of sources\cite{phinney91,narayan91}
suggest event rates of several per year and the radiation they
emit can be calculated unambiguously
\cite{peters63,blanchet89,lincoln90,damour91,wiseman91,wiseman92,kidder92}.
In support of the LIGO/VIRGO observational effort, theoreticians
must combine refined calculations of the radiation from binary
inspiral with the anticipated detector properties to deduce the
instrument sensitivity.  These calculations will, in turn, play
an important role in the final design and ultimate use of the
instruments. In this paper, we begin a detailed analysis of
binary inspiral as a source of gravitational radiation for a
LIGO/VIRGO-like interferometric detector.

Our goal in this work is to estimate the sensitivity of
LIGO/VIRGO-like interferometers and the LIGO detector by
determining the rate at which inspiralling binary systems can be
detected, the range to which they can be observed, and the
precision with which they can by characterized in a single
interferometer.  We also explore the compromises that must be
made as different questions are asked of the observations.  The
optimal design and operation of these interferometers depends on
a detailed understanding of the nature of the detector response
to the radiation, the detector noise power spectral density
(PSD), {\em and the questions the observation is meant to
resolve.\/} For example, we show in \S\ref{sec:implications} that
the goal of observing as many sources as possible (without
necessarily being able to characterize them precisely) leads to a
different optimal interferometer configuration than the goal of
characterizing observed sources as precisely as possible (while
allowing that weak sources may be missed entirely).

The ultimate goal of our assessment of binary inspiral is to
determine the ability of a single interferometer (or a network of
such interferometers) to
\begin{enumerate}
\item {\em Detect\/} the gravitational radiation from the last
few minutes of inspiral of a binary neutron star or black hole
system, and
\item {\em Measure\/} the parameters describing the detected
binary system from the observed gravitational radiation.
\end{enumerate}
By {\em detection\/} we mean the determination of the presence or
absence of a signal characteristic of an inspiralling binary
system in the output of a detector, {\em irrespective\/} of the
particular parameters that might characterize the observed binary
system. By {\em measurement} we mean the determination of the
parameters that characterize the signal {\em presumed to be
present\/} in the detector output.  In a real detector noise can
mask or distort a signal present in the detector output;
alternatively, it can conspire to appear as a signal
characteristic of a binary system.  Consequently, any conclusion
we draw from observations ({\em e.g.,\/} that we have detected an
inspiralling binary system) is associated with a probability that
characterizes our certainty in its validity.

As a practical matter, reliable detection of gravitational
radiation will initially require coincident observation between
two or more interferometers so that non-Gaussian noise events can
be differentiated from gravitational radiation signals. When
completed, the LIGO {\em detector\/} will consist of two {\em
interferometers:\/} one in Hanford, Washington and one in
Livingston, Louisiana\cite{abramovici92}. The relative
orientation of the two interferometers has been chosen to
maximize their sensitivity to a single polarization state of the
gravitational radiation impinging on the Earth\footnote{While
primarily sensitive to a single polarization state of the
radiation field, LIGO will, with more limited sensitivity, be
capable of observing both polarizations simultaneously. In a
subsequent paper we will take the exact detector orientations
into account in our analysis.}\cite{thorne92}; consequently, the
two interferometers act together like a single interferometer
whose sensitivity will be greater than that of either of its
components. This analogy is not exact: combining the
interferometers in this way ignores the different arrival times
of the signal at the two distinct interferometers. Nevertheless,
our single-detector analysis is directly relevant to the actual
LIGO configuration, and we discuss our results in the context of
{\em (i)\/} single LIGO/VIRGO-like {\em interferometers\/} of the
proposed initial and advanced designs, and {\em (ii)\/} the LIGO
{\em detector\/} in the limit that its two independent
interferometers share exactly the same orientation and ignoring
the difference in the signal arrival times.

In principle, simultaneous observation of an inspiralling binary
system in {\em three\/} interferometers of different orientations
is sufficient to measure (among other characteristics) the
binary's luminosity distance ${d_L}$, its position in the sky,
and a function ${\cal M}$ that depends only on the masses of its
components and its cosmological redshift\cite{schutz86,krolak87}.
If the position is known through other observations, then ${\cal
M}$ and ${d_L}$ can be measured from observations with only two
interferometers of different orientations. Observation of an
inspiralling binary in a single interferometer can measure ${\cal
M}$ and a waveform ``amplitude'' ${\cal A}$, which depends on the
luminosity distance and orientation of the binary with respect to
the detector.  Even this limited information is of astrophysical
significance, however: from observations of the distribution of
${\cal A}$ and ${\cal M}$ among the observed binaries one can
determine the distribution of the component masses of
inspiralling binary systems, find the Hubble constant and
deceleration parameter, and test cosmological
models\cite{chernoff92}.

The study of binary systems as sources of gravitational radiation
began in 1963 when Peters and Mathews\cite{peters63} made the
first detailed calculation of the gravitational radiation
luminosity from inspiralling binary systems, focusing on the
leading order quadrupole radiation from two point particles in
circular and elliptical orbits. Clark and Eardley\cite{clark77}
explored how gravitational radiation (among other effects) drives
the orbital evolution of binary systems. Clark\cite{clark79}
suggested inspiralling binary systems as an important source of
gravitational radiation for modern interferometric detectors;
however, he considered only the burst of radiation from the
coalescence event itself. Thorne recognized the importance of the
gravitational radiation from the final few minutes of {\em
inspiral.\/} Since 1987 a number of investigators have worked
with increasing sophistication on problems related to the
observation of inspiralling binaries in interferometric
detectors. Some have focused on understanding and refining
interferometer detector technology
\cite{meers88,vinet88,meers89,dhurandhar89a,saulson90,meers91},
others have focused on the data analysis problems of detecting or
determining the characteristics of a binary system from the
radiation
\cite{smith87,dhurandhar89b,lobo90,krolak91,sathyaprakash91,jaranowski92},
and still others have focused on refining our understanding of
the gravitational radiation waveform from these
systems
\cite{blanchet89,lincoln90,damour91,wiseman91,wiseman92,kidder92,kochanek92,bildsten92,junker92}
or their rate of occurrence in the
Universe\cite{phinney91,narayan91}.

The remainder of this paper is organized as follows. In
\S\ref{sec:review} we review how the precision with which the
parameterization of a signal, observed in a noisy detector, is
determined. In \S\ref{sec:application}, we apply these techniques
to the particular problem of finding the precision with which the
parameters of an inspiralling binary system can be measured.
These results depend only on the detector noise power spectral
density (PSD) and not on the type of detector ({\em e.g.,\/}
interferometer or bar). In \S\ref{sec:ligo} we apply the results
of \S\ref{sec:application} to both the initial and advanced
proposed LIGO/VIRGO-like interferometers.  We discuss the
astrophysical implications of these results in
\S\ref{sec:implications} and present our conclusions in
\S\ref{sec:conclusions}.


\section{Measurement and uncertainty}
\label{sec:review}

In this section we review the techniques we use to determine the
statistical uncertainties in observations of binary inspiral.  A
complete discussion is found in Finn\cite{finn92a}.  The
techniques developed there are closely related to those
associated with signal analysis by optimal filtering. For more
information on optimal filtering and signal analysis we direct
the reader to Oppenheim, Willsky, and Young\cite{oppenheim83},
the review by Davis\cite{davis89} and references therein,
Wainstein and Zubakov\cite{wainstein62}, and Hancock and
Wintz\cite{hancock66}.

\subsection{Introduction}
\label{subsec:measurement-intro}

Consider a deterministic ({\em i.e,\/} not stochastic) source of
gravitational radiation ({\em e.g.,\/} an inspiralling binary
system) and a detector ({\em e.g.,\/} a laser interferometer). We
write the response of the detector to the radiation as a
superposition of noise $n(t)$ and signal
$m(t;\bbox{\mu})$, where $\bbox{\mu}$ is a minimal
set of parameters that uniquely characterizes the detector
response (absent the noise) to the radiation for the entire
duration of the observation (in \S\ref{sec:application} we show
that for an inspiralling binary system observed by a single
interferometric detector $\bbox{\mu}$ is a four dimensional
vector). Let the source of radiation be characterized by
$\widetilde{\bbox{\mu}}$. In analyzing the output of
the detector, we have two goals:
\begin{enumerate}
\item To determine whether a signal is present in the detector
output, and
\item To find the precision with which we can determine
$\widetilde{\bbox{\mu}}$ (assuming a signal is
present).
\end{enumerate}

Owing to the detector noise, we cannot determine with certainty
either the presence of a signal or (assuming it present)
$\widetilde{\bbox{\mu}}$. Instead, we find the
probability that the detector output is consistent with the
presence of a signal, and represent our uncertainty in
$\widetilde{\bbox{\mu}}$ by a set of volumes $V(P)$ in
parameter space, such that $\widetilde{\bbox{\mu}}$ is
in $V(P)$ with probability $P$. The volumes $V(P)$ are a measure
of the sensitivity of the detector. Throughout this section, we
will assume that we have determined that a signal is present so
that the probability $P$ associated with $V(P)$ is a conditional
probability. In a later paper, we will discuss the determination
of the probability that a signal of the form
$m(t;\bbox{\mu})$ is present in the detector output.

Each observed signal is immersed in its own realization of the
detector noise; consequently we cannot know in advance of an
observation what the volumes $V(P)$ will look like. We can,
however, determine what the most likely volumes are for a given
observation, and that is what we do here. In this section we
review briefly the procedures used to find the most likely
volumes $V(P)$. These procedures are developed in
Finn\cite{finn92a}, and we refer the interested reader there for
more information.

\subsection{The probability volumes $V(P)$}
\label{subsec:V(P)}

We characterize the observed output of our detector as a time
series $g(t)$, which is a superposition of noise $n(t)$ and
(perhaps) a signal $m(t;\widetilde{\bbox{\mu}})$:
\begin{equation}
g(t)\equiv\left\{
{\arraycolsep=2\arraycolsep
\begin{array}{ll}
n(t)+m(t;\widetilde{\bbox{\mu}})&\text{signal
present}\\
n(t)&\text{signal absent.}
\end{array}}
\right.
\end{equation}
The parameter $\widetilde{\bbox{\mu}}$ is fixed but
unknown, and we assume the noise is drawn from a stationary
process.  The probability density that a signal with
parameterization $\bbox{\mu}$ is present in the detector
output $g(t)$ is
\begin{eqnarray}
P(\bbox{\mu}|g) &\equiv&
\left(
  \begin{tabular}{l}
The conditional probability that a\\
signal characterized by $\bbox{\mu}$ is present\\
given the detector output $g(t)$.
  \end{tabular}
\right)\nonumber\\
&=&{\Lambda(\bbox{\mu})\over\Lambda+P(0)/P(m)}
\label{eqn:P-signal}
\end{eqnarray}
where
\begin{mathletters}
\begin{eqnarray}
P(0) &\equiv& \left(
\begin{tabular}{l}
The {\em a priori\/} probability\\
that the signal is absent
\end{tabular}\right),\\
P(m) &\equiv& \left[
\begin{tabular}{l}
The {\em a priori\/} probability\\
that the signal $m(\bbox{\mu})$\\
(for undetermined $\bbox{\mu}$)\\
is present
\end{tabular}\right],\\
\Lambda &\equiv& \int d^N\mu\,\Lambda(\bbox{\mu}),
\label{eqn:Lambda}\\
\Lambda(\bbox{\mu}) &\equiv&
p(\bbox{\mu})\exp\left[
2\left<g,m(\bbox{\mu})\right>
-\left<m(\bbox{\mu}),m(\bbox{\mu})\right>
\right],\label{eqn:Lambda(mu)}\\
p(\bbox{\mu}) &\equiv&
\left(
  \begin{tabular}{l}
The {\em a priori\/} probability\\
density that the signal\\
parameterization is $\bbox{\mu}$.
  \end{tabular}
\right),\\
\left<r,s\right> &\equiv&
  \int_{-\infty}^\infty df\,
    {\widetilde{r}(f)\widetilde{s}^*(f)\over S_h(|f|)},\\
\widetilde{r}(f) &\equiv& \int_{-\infty}^\infty dt\, e^{2\pi i f
t}r(t),\\
S_h(f) &\equiv& \left[
\begin{tabular}{l}
One-sided detector noise\\
power spectral density (PSD)
\end{tabular}
\right].
\end{eqnarray}
\end{mathletters}

\narrowtext
The {\em likelihood ratio\/} $\Lambda$ (eqn.~\ref{eqn:Lambda}) is
proportional to the {\em a posteriori\/} probability that a
signal is present in the observed $g(t)$. When that probability
exceeds a given threshold we assume a signal is present and try
to determine $\widetilde{\bbox{\mu}}$. We characterize
an observation of $g(t)$ in terms of the mode ({\em i.e.,\/} the
maximum) of the distribution $P(\bbox{\mu}|g)$,
denoted $\widehat{\bbox{\mu}}$.  The mode of
$P(\bbox{\mu}|g)$ is also the mode of the {\em odds
ratio\/} $\Lambda(\bbox{\mu})$.  In terms of
$\widehat{\bbox{\mu}}$ the observation's
signal-to-noise ratio (SNR) $\rho^2$ is\footnote{Note that
$\rho^2$ is quadratic in the signal strength.  In the literature
SNR is often used to refer to both $\rho$ and $\rho^2$. We avoid
this ambiguity by using either $\rho$ or $\rho^2$ as
appropriate.}
\begin{equation}
\rho^2 =
2\left<m(\widehat{\bbox{\mu}}),
m(\widehat{\bbox{\mu}})\right> .
\label{defn:rho2}
\end{equation}
Assuming that $\Lambda(\bbox{\mu})$
is smooth and that its global maximum is a local extremum,
$\widehat{\bbox{\mu}}$ satisfies
\begin{eqnarray}
2\left<
m(\widetilde{\bbox{\mu}}) -
m(\widehat{\bbox{\mu}}),
{\partial m\over\partial\mu_j}
(\widehat{\bbox{\mu}})
\right>\nonumber\\
\qquad{}+{\partial\ln p\over\partial\mu_j}
(\widehat{\bbox{\mu}})
&=&
-2\left<
n,{\partial m\over\partial\mu_j}
(\widehat{\bbox{\mu}})
\right>.
\label{eqn:linearize-me}
\end{eqnarray}
We assume that the noise $n(t)$ is a normal random variable with
zero mean; consequently, so are each of the $\left<n,\partial
m/\partial\mu_j\right>$ on the righthand side of
equation~\ref{eqn:linearize-me}. Denote these random variables
$\nu_i$:
\begin{equation}
\nu_i \equiv
2\left<n,
{\partial m\over\partial\mu_i}(\widehat{\bbox{\mu}})
\right> .
\end{equation}
Since the $\nu_i$ are normal, their joint distribution is a
multivariate Gaussian, characterized by the means
$\overline{\nu_i}$, which vanish, and the quadratic moments
\begin{mathletters}
\begin{eqnarray}
\overline{\nu_i\nu_j} &=& 4\overline{
\left<n,{\partial m\over\partial\mu_i}
(\widehat{\bbox{\mu}})\right>
\left<n,{\partial m\over\partial\mu_j}
(\widehat{\bbox{\mu}})\right>
}\nonumber\\
&=&
2\left<
{\partial m\over\partial\mu_i}
(\widehat{\bbox{\mu}}),
{\partial m\over\partial\mu_j}
(\widehat{\bbox{\mu}})
\right>\\
&\equiv&{\cal C}^{-1}_{ij}.\label{defn:cij}
\end{eqnarray}
\end{mathletters}
Here we have used an overbar to indicate an average over all
instances of the noise $n(t)$. Since we have assumed that the
noise is normal these averages can be evaluated using the ergodic
theorem ({\em cf.\/}~Finn\cite{finn92a} for more details).  In
terms of ${\cal C}_{ij}$ ({\em i.e.,\/} the inverse of ${\cal
C}^{-1}_{ij}$), the joint distribution of the $\nu_i$ is given by
the probability density
\begin{equation}
p(\bbox{\nu}) =
{\exp\left[-{1\over2}\sum_{i,j}{\cal C}_{ij}\nu_i\nu_j\right]\over
\left[\left(2\pi\right)^N\det||{\cal C}^{-1}_{ij}||\right]^{1/2}
}\label{eqn:mv-gaussian}
\end{equation}
This is also the joint distribution of the quantities that appear
on the lefthand side of equation~\ref{eqn:linearize-me}. In
Finn\cite{finn92a} it was stated that for an observation
characterized by $\widehat{\bbox{\mu}}$, the
probability volumes $V(P)$ are given implicitly by
\begin{eqnarray}
K^2&\geq&
\sum_{i,j}{\cal C}_{ij}\left[
2\left<
m({\bbox{\mu}}) -
m(\widehat{\bbox{\mu}}),
{\partial m\over\partial\mu_i}(\widehat{\bbox{\mu}})
\right> +
{\partial\ln p\over\partial\mu_i}
(\widehat{\bbox{\mu}})
\right]\nonumber\\
&&\quad{}\times
\left[
2\left<
m({\bbox{\mu}})
- m(\widehat{\bbox{\mu}}),
{\partial m\over\partial\mu_j}(\widehat{\bbox{\mu}})
\right> +
{\partial\ln p\over\partial\mu_j}
(\widehat{\bbox{\mu}})
\right],\nonumber\\
&&\label{eqn:exact-dist}
\end{eqnarray}
where
\begin{equation}
P = \int_{\sum_{i,j}{\cal C}_{ij}\nu_i\nu_j\leq K^2} d^N\nu
{\exp\left[-{1\over2}\sum_{i,j}{\cal C}_{ij}\nu_i\nu_j\right]\over
\left[\left(2\pi\right)^N\det||{\cal C}^{-1}_{ij}||\right]^{1/2}
}\label{eqn:exact-dist-P}
\end{equation}
and $\widetilde{\bbox{\mu}}\in V(P)$ with probability
$P$ (recall that $P$ is conditional on the assumption that a
signal is in fact present in the detector output). {\em This is
correct only when $\Lambda(\bbox{\mu})$ has a single
extremum, or near $\widehat{\bbox{\mu}}$.} In the most
general case only a Monte Carlo analysis can determine the
probability volumes $V(P)$. No other results of
Finn\cite{finn92a} are affected by this correction.

\subsection{The strong signal approximation}
\label{subsec:strong-signal}

The volumes $V(P)$ are representations of a cumulative
probability distribution function. Denote the corresponding
probability density by
$P(\delta\bbox{\mu}|\widehat{\bbox{\mu}})$:
\begin{equation}
P(\delta\bbox{\mu}|\widehat{\bbox{\mu}})
\equiv\left(
\begin{tabular}{l}
The conditional probability\\
density that $m$ is characterized\\
by $\widehat{\bbox{\mu}}+\delta\bbox{\mu}$,
given that $\Lambda(\bbox{\mu})$\\
has mode $\widehat{\bbox{\mu}}$
\end{tabular}
\right)
\end{equation}
In the limit of large $\rho^2$, $P(\delta\bbox{\mu}|
\widehat{\bbox{\mu}})$ becomes sharply peaked about
$\widehat{\bbox{\mu}}$ and the determination of $V(P)$
is greatly simplified.  Suppose that $\rho^2$ is so large that
for $\widetilde{\bbox{\mu}}\in V(P)$ for all $P$ of
interest the difference $m(\widetilde{\bbox{\mu}})-
m(\widehat{\bbox{\mu}})$ can be linearized in
$\delta\bbox{\mu}$, where
\begin{equation}
\delta\bbox{\mu} \equiv
\widetilde{\bbox{\mu}} -
\widehat{\bbox{\mu}} .
\end{equation}
We then obtain in place of
equation~\ref{eqn:linearize-me}
\begin{equation}
\sum_{i,j}\delta\mu_i\,{\cal C}_{ij}^{-1}
= -2\left<n,
{\partial m\over\partial\mu_j}(\widehat{\bbox{\mu}})
\right> -
{\partial\ln
p\over\partial\mu_j}(\widehat{\bbox{\mu}})
\label{eqn:delta-mu-j}
\end{equation}
The random variables $\delta\bbox{\mu}$ are related to
the $\bbox{\nu}$ by a linear transformation,
\begin{equation}
\delta\mu_i =
-\sum_{i,j}{\cal C}_{ij}\left[\nu_j+
{\partial\ln p\over\partial\mu_j}(\widehat{\bbox{\mu}})
\right];
\end{equation}
consequently, the $\delta\bbox{\mu}$ are normal with
means
\begin{equation}
\overline{\delta\mu_i} =
-\sum_{i,j}{\cal C}_{ij}{\partial\ln p\over\partial\mu_j}
(\widehat{\bbox{\mu}}),
\end{equation}
and quadratic moments
\begin{eqnarray}
\overline{
\left(\delta\mu_i-\overline{\delta\mu_i}\right)
\left(\delta\mu_j-\overline{\delta\mu_j}\right)}
= {\cal C}_{ij}.
\end{eqnarray}
The probability distribution
$P(\delta\bbox{\mu}|\widehat{\bbox{\mu}})$
is thus a multivariate Gaussian ({\em cf.\/}
eqn.~\ref{eqn:mv-gaussian}):
\begin{equation}
P(\delta\bbox{\mu}|\widehat{\bbox{\mu}}) =
{\exp\left[
-{1\over2}\sum_{i,j}{\cal C}^{-1}_{ij}
\left(\delta\mu_i-\overline{\delta\mu_i}\right)
\left(\delta\mu_j-\overline{\delta\mu_j}\right)
\right]\over
\left[\left(2\pi\right)^N\det||{\cal C}_{ij}||\right]^{1/2}}.
\label{eqn:prob-dist}
\end{equation}
Note that the matrix ${\cal C}_{ij}$ has now acquired a physical
meaning: in particular, we see that the variances $\sigma^2_i$ of
the $\delta\mu_i$ are
\begin{eqnarray}
\sigma^2_i &\equiv& \overline{
\left(\delta\mu_i-\overline{\delta\mu_i}\right)^2}\nonumber\\
&=& {\cal C}_{ii}\label{defn:sigma}
\end{eqnarray}
and the correlation coefficients $r_{ij}$ are given by
\begin{eqnarray}
r_{ij} &\equiv&\sigma_i^{-1}\sigma_j^{-1}
\overline{
\left(\delta\mu_i-\overline{\delta\mu_i}\right)
\left(\delta\mu_j-\overline{\delta\mu_j}\right)}
\nonumber\\
&=&{{\cal C}_{ij}\over\sigma_i\sigma_j}
\label{defn:correlation}
\end{eqnarray}
In this sense we say that ${\cal C}_{ij}$ is the covariance
matrix of the random variables $\delta\bbox{\mu}$.

In the strong signal approximation, the surfaces bounding the
volume $V(P)$ are ellipsoids defined by the equation
\begin{equation}
\sum_{i,j}\left(\delta\mu_i-\overline{\delta\mu_i}\right)
\left(\delta\mu_j-\overline{\delta\mu_j}\right)
{\cal C}^{-1}_{ij} = K^2,
\end{equation}
where the constant $K^2$ is related to $P$ by
\begin{equation}
P=\int_{\sum_{i,j}{\cal C}^{-1}_{ij}x^ix^j\leq K^2}
d^N\! x\,
{\exp\left[-{1\over2}\sum_{i,j}{\cal C}^{-1}_{ij}x^ix^j\right]
\over\left[\left(2\pi\right)^N\det||{\cal C}_{ij}||\right]^{1/2}}
\end{equation}

Finally we come to the question of when the linearization in
equation \ref{eqn:delta-mu-j} is a reasonable approximation.
Three considerations enter here:
\begin{enumerate}
\item The probability contours of interest ({\em e.g.,\/} 90\%)
must not involve $\delta\bbox{\mu}$ so large that the
linearization of $m(\widetilde{\bbox{\mu}}) -
m(\widehat{\bbox{\mu}})$ is a poor approximation;
\item The probability contours of interest
must not involve $\delta\bbox{\mu}$ so large that for
$\delta\bbox{\mu}\in V(P)$,
$\Lambda(\delta\bbox{\mu} +
\widehat{\bbox{\mu}})$
has more than one extremum or inflection point; and
\item The condition number of the matrix ${\cal C}^{-1}_{ij}$
must be sufficiently small that the inverse ${\cal C}_{ij}$ is
insensitive to the linearization approximation in the
neighborhood of
$\widehat{\bbox{\mu}}$.\footnote{Recall that the
relative error in $\delta\bbox{\mu}$ is the condition
number times the relative error in ${\cal C}_{ij}^{-1}$: for a
large condition number, small errors in ${\cal C}_{ij}^{-1}$
introduced by the linearization approximation can result in large
errors in $\delta\bbox{\mu}$ ({\em cf.\/} Golub \& Van
Loan\cite{golub89}).}
\end{enumerate}
If the validity of the linearization procedure for a particular
problem is doubtful owing to the violation of any of these
conditions, then we must fall-back on either
equations~\ref{eqn:linearize-me}, \ref{eqn:exact-dist} and
\ref{eqn:exact-dist-P} (if $\Lambda(\bbox{\mu})$ has a
single extremum), or a Monte Carlo analysis.

\section{Precision of measurement:\protect{\\}
Binary inspiral}
\label{sec:application}

In this section we apply the techniques described in
\S\ref{sec:review} to the problem of measuring the
characteristics of an inspiralling binary system in an
interferometric gravitational radiation detector. We determine
the gravitational radiation from an evolving binary system in the
quadrupole approximation and evaluate the corresponding detector
response.  The amplitude of the response is a function of the
(unknown and unknowable) relative orientation of the source and
the detector, and we evaluate its mean square amplitude and
{\em a priori\/} cumulative probability distribution.  We also
discuss the validity of describing binary evolution using the
quadrupole approximation, arguing that while the result is
certainly not good enough for use in actual data analysis, it is
sufficient for the purpose of exploring the precision with which
we will ultimately be able to characterize a binary system
through gravitational radiation observations in LIGO/VIRGO-like
interferometers.

The general interferometer response to the gravitational
radiation from an inspiralling binary system is a sinusoid with
slowly varying amplitude and frequency. Using the stationary
phase approximation, we obtain an analytic expression for the
Fourier transform of the response. We find that the SNR $\rho^2$
and the covariance matrix ${\cal C}_{ij}$ can be expressed simply
in terms of several moments of $S^{-1}_h(f)$, the inverse of the
interferometer noise PSD.

\subsection{Parameterization of the radiation waveform}
\label{subsec:parameterization}

In order to express the response of a single interferometer to
the gravitational radiation from a binary system (or any source),
we define two coordinate systems: the {\em source\/} coordinate
system and the {\em radiation\/} coordinate system.  The binary
system is most simply described in the source coordinate system.
The ${\bf e}^S_z$-axis of this coordinate system is along the
binary system's angular momentum.  We choose the axes ${\bf
e}^S_x$ and ${\bf e}^S_y$ of the source coordinate system to make
the expression of the radiation directed toward the
interferometer simple: the ${\bf e}^S_x$ axis is chosen so that
the unit vector ${\bf n}_I$ in the direction of the
interferometer is in the ${\bf e}^S_x\wedge{\bf e}^S_z$ plane and
in the positive ${\bf e}^S_x$ direction (if the interferometer is
along the polar axis, then there is no preferred direction for
the ${\bf e}_x^S$ axis).

The radiation coordinate system has its ${\bf e}^R_z$ axis in the
direction ${\bf n}_I$, and its ${\bf e}^R_x$ and ${\bf e}^R_y$
axes are projections of the ${\bf e}^S_x$ and ${\bf e}^S_y$
coordinate axes normal to ${\bf n}_I$:
\begin{mathletters}
\begin{eqnarray}
{\bf e}^R_y &=& {\bf e}^S_y,\label{defn:ery}\\
{\bf e}^R_x &=& {\bf e}^S_y\times{\bf n}_I.\label{defn:erx}
\end{eqnarray}
\end{mathletters}
In the radiation coordinate system, the radiation propagating
toward the interferometer is described by
\begin{equation}
{\bf h} = {h_{{}+{}}}{\bf e}_{{}+{}}+{h_{{}\times{}}}{\bf
e}_\times
\end{equation}
where
\begin{mathletters}
\begin{eqnarray}
{\bf e}_{{}+{}} &=&
{\bf e}^R_x\otimes{\bf e}^R_x -
{\bf e}^R_y\otimes{\bf e}^R_y\\
{\bf e}_{\times} &=&
{\bf e}^R_x\otimes{\bf e}^R_y +
{\bf e}^R_y\otimes{\bf e}^R_x.
\end{eqnarray}
\end{mathletters}

\subsubsection{The quadrupole waveform}
\label{subsubsec:quadrupole-waveform}

In the quadrupole approximation\footnote{Here and henceforth we
adopt units where $G=c=1$},
\begin{mathletters}
\begin{eqnarray}
{h_{{}+{}}}&\equiv&
2{{\cal M}\over{d_L}}\left(1+\cos^2 i\right)\left(\pi{\cal M}
f\right)^{2/3}
\cos\left(\Phi+\Psi\right)
\label{defn:hplus}\\
{h_{{}\times{}}}&\equiv&
4{{\cal M}\over{d_L}}\cos i\left(\pi{\cal M} f\right)^{2/3}
\sin\left(\Phi+\Psi\right)
\label{defn:hcross}
\end{eqnarray}
\end{mathletters}
\begin{mathletters}
where
\begin{eqnarray}
{d_L}&\equiv&\left(
\begin{tabular}{l}
Luminosity distance to binary
\end{tabular}
\right),\\
\cos i &\equiv& {\bf n}_I\cdot{\bf e}^S_x,\\
i&\equiv&\left(
\begin{tabular}{l}
Inclination angle of orbital\\
angular momentum to line of\\
sight toward the interferometer.
\end{tabular}
\right),\\
{\cal M} &\equiv&\left(\text{Chirp mass}\right),\nonumber\\
&=&\left(1+z\right)\mu^{3/5}M^{2/5},\label{defn:M}\\
\Phi &\equiv& -2\left(T-t\over5{\cal M}\right)^{5/8},
\label{defn:Phi}\\
f &\equiv& {1\over2\pi}{\partial\Phi\over\partial t},\nonumber\\
&=&{1\over\pi{\cal M}}\left[{5\over256}{{\cal M}\over
T-t}\right]^{3/8},
\label{defn:f}
\end{eqnarray}
\end{mathletters}
$\Psi$ is the phase of the binary system at $t=T$, $T$ is the
Newtonian ``moment of coalescence,'' $z$ is the cosmological
redshift of the binary system, and $M$ and $\mu$ are the binary
system's total and reduced mass\cite{peters63,thorne87}.  The
gravitational radiation frequency $f$ is twice the systems
orbital frequency.

\narrowtext
\subsubsection{The interferometer response}

The response of an interferometer to the gravitational radiation
field is a linear combination of ${h_{{}+{}}}$ and
${h_{{}\times{}}}$:
\begin{equation}
m = F_{{}+{}}h_{{}+{}} + F_\times h_\times,
\end{equation}
where the {\em antenna patterns\/} $F_{{}+{}}$ and $F_\times$
depend on the orientation of the interferometer with respect to
the binary system\cite{thorne87}. We make no assumptions
regarding the relative orientation of the interferometer and the
binary; consequently, the general interferometer response is
\begin{equation}
m(t;{\cal A},{\cal M},\psi,T) \equiv
{\cal A}{\cal M}\left(\pi
f{\cal M}\right)^{2/3}\cos\left(\Phi+\psi\right),
\label{defn:m}
\end{equation}
where $\psi$ is a constant (distinct from $\Psi$ in
equations~\ref{defn:hplus} and \ref{defn:hcross}), and $\Phi$ and
$f$ are given above in equations \ref{defn:Phi} and \ref{defn:f}.
We will return in \S\ref{subsec:prop-Theta} to discuss how ${\cal
A}$ depends on the orientation angles through $F_{{}+{}}$ and
$F_\times$.

\subsubsection{Radiation reaction and the quadrupole
approximation}
\label{subsubsec:back-reaction}

LIGO/VIRGO-like interferometers are most sensitive to
gravitational radiation with frequencies in the range
30--1000~Hz. This corresponds to binary orbital frequencies of
15--500~Hz. In this regime gravitational radiation reaction is
the most important factor in determining the evolution of a
binary system's orbit\cite{kochanek92,bildsten92}. In
\S\ref{subsubsec:quadrupole-waveform}, where we describe the
evolution of the binary systems orbit, only backreaction owing to
the leading-order quadrupole radiation is taken into
account\cite{peters63,thorne87}.  This approximation neglects
higher order effects (in both $v/c$ and $M/r$) that contribute to
the gravitational radiation luminosity and, consequently, the
evolution of $f$. This has serious ramifications for the
construction of the model detector response\cite{cutler92b}.

The detector response to binary inspiral is a sinusoid of slowly
varying amplitude and frequency. The determination of the
characteristics of the binary system is equivalent to finding the
``template'' response that is most closely correlated to the
detector response. If the phase of the template drifts from that
of the signal by as little as $\pi$ radians over the course of
the observations, then the correlation will be insignificant.
Neutron star binary inspiral observations in LIGO/VIRGO-like
interferometers will last for on order $2\pi\times10^4$~radians
in phase; consequently, the phase advance can be determined to
better than 1~part in~$10^4$. The errors we have made in our
template $m(\bbox{\mu})$ (eqn.~\ref{defn:m}) by
neglecting the post-Newtonian contributions to the evolution of
the binary system lead to phase differences significantly greater
than $2\pi$~radians over $10^4$~cycles observed. Consequently,
the waveform model used in actual data analysis must be more
accurate than that given by the quadrupole
approximation\cite{cutler92a,cutler92b}\footnote{Cutler, Finn,
Poisson, and Sussman\cite{cutler92a} have shown that successive
post-Newtonian approximations to the evolution and waveform
converge very slowly upon the fully relativistic solution.
Consequently, it may require an impractically high order
post-Newtonian expansion to predict correctly the advance of the
phase over the course of the LIGO/VIRGO observations.}.

Nevertheless, for the particular purpose of exploring our ability
to detect and characterize binary systems by their gravitational
radiation signature, we believe the quadrupole approximation
waveform is a useful substitute for a more accurate waveform.
The determination of the {\em anticipated\/} sensitivity of a
LIGO/VIRGO-like interferometer to binary inspiral depends on the
$\rho^2$ and ${\cal C}_{ij}$, and we expect that the quadrupole
approximation provides good lower bounds on these. The
predominant observable effect of the inclusion of post-Newtonian
corrections is to change the rate that the wave frequency $f$
advances, thus changing the elapsed phase of the wave over the
period of the observation. The SNR $\rho^2$ depends approximately
on the elapsed phase and the corrections, while large compared to
$2\pi$, change this by only a small fraction of the total.
Similarly, in the quadrupole approximation the rate at which $f$
advances depends exclusively on ${\cal M}$ ({\em
cf.\/}~eqn.~\ref{defn:f}); consequently, to the extent that the
corrections to the quadrupole formula depend on characteristics
of the binary other than ${\cal M}$ ({\em e.g.,\/} component
masses and spins) these characteristics are observable and affect
the precision with which ${\cal M}$ can be measured. In this way
we expect that the use of a more accurate waveform in our
analysis will increase the estimated $\sigma_{\cal M}$ but have
little effect on our estimates of $\sigma_{\cal A}$,
$\sigma_\psi$, and $\sigma_T$ made using the quadrupole waveform
and corresponding binary evolution.

\subsection{The stationary phase approximation}
\label{subsec:stationary-phase}

In order to evaluate the SNR $\rho^2$ (eqn.~\ref{defn:rho2}) and
the covariance matrix ${\cal C}_{ij}$ (eqn.~\ref{defn:cij}) we
must find the Fourier transform $\widetilde{m}$ of $m$
(eqn.~\ref{defn:m}). We approximate $\widetilde{m}$ using the
method of stationary phase.  Given a real function of the form
\begin{equation}
k(t) = A(t)\cos\widehat\Phi(t),
\end{equation}
where $\partial\widehat\Phi/\partial t$ is a monotonically
increasing function of $t$, the stationary phase approximation to
the Fourier transform $\widetilde{k}(f)$ is
\widetext
\begin{mathletters}
\begin{eqnarray}
\widetilde{k}(f) &=& \int_{-\infty}^\infty dt\,k(t)e^{2\pi ift}\\
&\simeq&
\left\{
  \begin{array}{ll}
  {1\over2}A\left[{\cal T}(f)\right]
  \left[2\pi/{\partial^2\widehat\Phi\over\partial
  t^2}[{\cal T}(f)]\right]^{1/2}\\
  \qquad\times\exp\left[i
  \left(2\pi f{\cal T}(f)-\widehat\Phi({\cal T}(f))+\pi/4\right)
  \right]&\qquad\text{for $f>0$}\\
  \widetilde{k}^*(-f)&\qquad\text{for $f<0$},
  \end{array}\right.
\nonumber\\
\label{defn:tilde-m}
\end{eqnarray}
\end{mathletters}
where
\begin{equation}
{\cal T}\left[{1\over2\pi}{\partial\widehat\Phi\over\partial
t}(t)\right] = t.
\end{equation}

\narrowtext
The validity of the approximation rests on the assumption that
the amplitude $A$ and the angular frequency
$\partial\widehat\Phi/\partial t$ change slowly over a period:
\begin{mathletters}
\begin{eqnarray}
{\partial\ln A/\partial t\over\partial\widehat\Phi/\partial t}
&\ll&
1\qquad\text{and}\\
{\partial^2\widehat\Phi/\partial t^2\over
\left(\partial\widehat\Phi/\partial t\right)^2} &\ll& 1.
\end{eqnarray}
\end{mathletters}
For the interferometer response $m$ given in
equation~\ref{defn:m} these two relations are equivalent to the
single condition
\begin{equation}
{T-t\over{\cal M}} \gg
\left({3\pi\over8}\right)^{8/5}\left(256\over5\right)^{3/5}
\simeq14,
\end{equation}
or, alternatively,
\begin{equation}
f{\cal M} \ll
{1\over\pi}
\left({5\over256}{8\over3\pi}\right)^{3/5}\simeq{1\over 37} .
\end{equation}
For binary systems that will be observable by LIGO/VIRGO-like
interferometers,
\begin{equation}
f{\cal M} = 6\times10^{-4}{f\over100\,{\rm Hz}}{{\cal
M}\over1.2\,{\rm M}_\odot}
\ll{1\over37};
\end{equation}
consequently, the stationary phase approximation is a good one
for our purposes.  We thus have
\widetext
\begin{equation}
\widetilde{m}(f) \simeq \left\{
{\arraycolsep=2\arraycolsep\begin{array}{ll}
{\cal A}{\cal M}^2
\left[{5\pi\over384}\right]^{1/2}\left(\pi f{\cal
M}\right)^{-7/6}\\
\qquad\times\exp\left\{
i\left[2\pi fT +{3\over128}\left(\pi f{\cal M}\right)^{-5/3}
- \psi + {\pi\over4}
\right]\right\}&\qquad\text{for $f>0$,}\\
\widetilde{m}^*(-f)&\qquad\text{for $f<0$.}
\end{array}}
\right.
\end{equation}

\narrowtext
\subsection{The signal-to-noise ratio}

Now suppose that we have analyzed the output $g(t)$ of an
interferometer for the signal $m$ and found that the likelihood
function is maximized for the parameterization $\{\widehat{\cal
A}$, $\widehat{\cal M}$, $\widehat\psi$, $\widehat{T}\}$. The SNR
$\rho^2$ is then given by ({\em cf.\/} eqn.~\ref{defn:rho2})
\begin{eqnarray}
\rho^2 &=&
2\left<m(\widehat{\bbox{\mu}}),
m(\widehat{\bbox{\mu}})\right>\nonumber\\
&=&{5\over96\pi^{4/3}}\widehat{\cal A}^2\widehat{\cal
M}^{5/3}f_{7/3},
\label{eqn:binary-snr}
\end{eqnarray}
where, if we assume that we have access to $g(t)$ for all $t$,
\begin{equation}
f_{7/3} = \int^\infty_0 df\,\left[{f^{7/3} S_h(f)}\right]^{-1}.
\label{defn:infinite-f7/3}
\end{equation}

In practice, of course, data analysis is limited to a finite
length sample of the interferometer output. During this limited
interval, the signal ``frequency'' $f$ (eqn.~\ref{defn:f}) of an
inspiralling binary system ranges from $f_l$ to $f_h$. We assume
that the minimum of $S_h(f)$ occurs in this interval and not too
near the endpoints. Then, just as the slowly varying amplitude
and frequency of $m$ permitted us to estimate $\widetilde{m}(f)$
using the stationary phase approximation, so it allows us to
approximate $\rho^2$ for a finite duration observation by
replacing the lower and upper limits in the integral expression
for $f_{7/3}$ (eqn.~\ref{defn:f7/3}) by $f_l$ and $f_h$. The
lower bound $f_l$ is determined by the duration of the data
stream being analyzed while the upper bound $f_h$ is determined
by the coalescence of the binary components.  We further assume
that the $f_{7/3}$ does not change significantly as
$f_h\rightarrow\infty$; thus, we have
\begin{equation}
f_{7/3} = \int^\infty_{f_l} df\,\left[{f^{7/3} S_h(f)}\right]^{-1}.
\label{defn:f7/3}
\end{equation}
We discuss our choice of low-frequency cut-off $f_l$ in
\S\ref{subsubsec:cutoff} and again in \S\ref{subsec:cut-off}.

\subsection{The covariance matrix}

Turn now to the calculation of the covariance matrix ${\cal
C}_{ij}$ ({\em cf.\/}~eqn.~\ref{defn:cij}). Instead of a
parameterization in terms of ${\cal A}$ and ${\cal M}$, it is
more convenient to introduce $\eta$ and $\zeta$ defined by
\begin{mathletters}
\begin{eqnarray}
\widehat{{\cal A}}(1+\eta) &\equiv& {\cal A}\\
\widehat{{\cal M}}(1+\zeta) &\equiv& {\cal M},
\end{eqnarray}
\end{mathletters}
where $\widehat{{\cal A}}$ and $\widehat{{\cal M}}$ are the modes
of the observed distribution of ${\cal A}$ and ${\cal M}$.

\widetext
Given our expression for $\widetilde{m}(f)$, we can evaluate all
of the elements of the symmetric matrix ${\cal C}^{-1}_{ij}$ in
terms of the {\em frequency moments\/} $\overline{f}_\beta$
defined by
\begin{equation}
\overline{f}_\beta \equiv f^{-1}_{7/3}\int_{f_l}^\infty
df\,\left[{f^{\beta} S_h(f)}\right]^{-1}.\label{defn:f-beta}
\end{equation}
To express ${\cal C}_{ij}$, only $\overline{f}_\beta$ for
$\beta\in\left\{17/3,\, 4,\, 3,\, 7/3,\, 1/3\right\}$ are
needed. In terms of these moments,
\begin{equation}
{\cal C}^{-1}_{ij} = \rho^2\left(
{\arraycolsep=4\arraycolsep
\begin{array}{cccc}
1&{5\over6}&0&0\\
&{25\over36}\left[
  1+{9\over4096}
  {\overline{f}_{17/3}\over
    \left(\pi\widehat{{\cal M}}\right)^{10/3}}
  \right]&
{5\over128}
{\overline{f}_{4}\over\left(\pi\widehat{{\cal M}}\right)^{5/3}}&
-{5\pi\over64}
{\overline{f}_3\over\left(\pi\widehat{{\cal M}}\right)^{5/3}}\\
&&1&-2\pi\overline{f}_{4/3}\\
&&&\left(2\pi\right)^2\overline{f}_{1/3}
\end{array}}
\right),\label{defn:invcij}
\end{equation}
where the indices are ordered $\eta$, $\zeta$, $\psi$, and $T$.
We have inverted equation~\ref{defn:invcij} to find ${\cal
C}_{ij}$ and so the variances and correlation coefficients
describing the distributions of $\eta$, $\zeta$, $\psi$, and $T$
({\em cf.\/} eqns.~\ref{defn:sigma} and~\ref{defn:correlation}).
The variances are
\begin{mathletters}
\begin{eqnarray}
\sigma^2_\eta &=& {\sigma^2_{\cal A}\over\widehat{{\cal
A}^2}}\nonumber\\
&=&
\left[
  1 + {4096\over9\Delta}
  \left(\pi\widehat{{\cal M}}\right)^{10/3}
  \left(\overline{f}_{1/3}-\overline{f}_{4/3}^2\right)
\right]\rho^{-2}
\label{eqn:sigma-eta}\label{eqn:sigma-A}\\
\sigma^2_\zeta &=&
{\sigma^2_{\cal M}\over\widehat{{\cal M}^2}}\nonumber\\
&=&
{16384\over25\Delta\rho^2}\left(
\overline{f}_{1/3}-\overline{f}_{4/3}^2
\right)\left(\pi\widehat{{\cal M}}\right)^{10/3}
\label{eqn:sigma_zeta}\label{eqn:sigma-M}\\
\sigma^2_\psi &=&
{\overline{f}_{17/3}\overline{f}_{1/3} -
\overline{f}^2_{3}\over\Delta\rho^2}
\label{eqn:sigma-psi}\\
\sigma^2_T &=&
{\overline{f}_{17/3}-\overline{f}_{4}^2
\over4\pi^2\Delta\rho^2}
\label{eqn:sigma-T}
\end{eqnarray}
\end{mathletters}
and the correlation coefficients are
\begin{mathletters}
\begin{eqnarray}
r_{\eta\zeta} &=& r_{{\cal A}{\cal M}}\nonumber\\
&=&
{64\left(\overline{f}^2_{4/3}-\overline{f}_{1/3}\right)
\left(\pi\widehat{\cal M}\right)^{10/3}
\over
\left\{
\left[
9\Delta+
4096\left(\pi\widehat{\cal M}\right)^{10/3}
\left(\overline{f}_{1/3}-\overline{f}^2_{4/3}\right)
\right]
\left(\overline{f}_{1/3}-\overline{f}^2_{4/3}\right)
\left(\pi\widehat{\cal M}\right)^{10/3}
\right\}^{1/2}
},
\label{eqn:r-A-M}\\
r_{\eta\psi} &=& r_{{\cal A}\psi}\nonumber\\
&=&
{64\left(
  \overline{f}_{1/3}\overline{f}_4 -
  \overline{f}_{4/3}\overline{f}_{3}
\right)
\left(\pi\widehat{{\cal M}}\right)^{5/3}
\over
\left\{
\left[
9\Delta+
4096\left(\pi\widehat{\cal M}\right)^{10/3}
\left(\overline{f}_{1/3}-\overline{f}^2_{4/3}\right)
\right]
\left(\overline{f}_{17/3}\overline{f}_{1/3} -
\overline{f}_3^2\right) \right\}^{1/2}},
\label{eqn:r-A-psi}\\
r_{\eta T} &=& r_{{\cal A} T}\nonumber\\
&=&
{64
\left(
  \overline{f}_4\overline{f}_{4/3} -
  \overline{f}_3
\right)
\left(\pi\widehat{{\cal M}}\right)^{5/3}
\over
\left\{
\left[
9\Delta+
4096\left(\pi\widehat{\cal M}\right)^{10/3}
\left(\overline{f}_{1/3}-\overline{f}^2_{4/3}\right)
\right]
\left(\overline{f}_{17/3}-\overline{f}_{4}^2\right)
\right\}^{1/2}
},
\label{eqn:r-A-T}\\
r_{\zeta\psi} &=& r_{{\cal M}\psi}\nonumber\\
&=&
{
\left(
 \overline{f}_{4/3}\overline{f}_{3} -
 \overline{f}_{1/3}\overline{f}_{4}
\right)\over
\left[
\left(\overline{f}_{1/3}-\overline{f}_{4/3}^2\right)
\left(\overline{f}_{17/3}\overline{f}_{1/3} -
\overline{f}_3^2\right) \right]^{1/2}
},\label{eqn:r-M-psi}\\
r_{\zeta T} &=& r_{{\cal M} T}\nonumber\\
&=&
{
\overline{f}_3 -
\overline{f}_{4}\overline{f}_{4/3}
\over
\left[
\left(\overline{f}_{1/3}-\overline{f}_{4/3}^2\right)
\left(\overline{f}_{17/3}-\overline{f}_{4}^2\right)
\right]^{1/2}
},
\label{eqn:r-M-T}\\
r_{\psi T}
&=&
{
\overline{f}_{17/3}\overline{f}_{4/3} -
\overline{f}_4\overline{f}_3
\over
\left[
\left(\overline{f}_{17/3}\overline{f}_{1/3} -
\overline{f}^2_{3}\right) \left(\overline{f}_{17/3} -
\overline{f}_{4}^2\right)
\right]^{1/2}
},
\label{eqn:r-psi-T}
\end{eqnarray}
\end{mathletters}
where
\begin{equation}
\Delta \equiv
\left(\overline{f}_{1/3} -
\overline{f}^2_{4/3}\right)\overline{f}_{17/3} -
\overline{f}_{1/3}\overline{f}_{4}^2
+\left(
  2\overline{f}_4\overline{f}_{4/3} -
  \overline{f}_3
\right)\overline{f}_3 > 0.
\end{equation}

\narrowtext
\subsection{Properties of ${\cal A}$}
\label{subsec:prop-Theta}

Despite the fact that we cannot measure the orientation angles
relating the interferometer to the source, we still need them in
order to assess the interferometer's sensitivity.  Recalling that
\begin{equation}
m = F_{{}+{}}h_{{}+{}} + F_\times h_\times,
\end{equation}
we express $F_{{}+{}}$ and $F_\times$ according to the following
convention:
\begin{enumerate}
\item Assume that the interferometer arms are the same length and
that they meet in right angles.
\item Define a right-handed coordinate system with one
interferometer arm along the $x$-axis and the other along the
$y$-axis.  Denote the unit vector in the direction of the $x$ arm
by ${\bf l}$ and the unit vector in the direction of the $y$ arm
by ${\bf m}$.
\item Let the position of a source in the sky be given by the
polar angle $\theta$ and the azimuthal angle $\phi$, and denote
the unit vector pointing toward the source by ${\bf n}_S$ ({\em
i.e.,\/} $-{\bf n}_I$).
\item The interferometer responds linearly to the radiation field
${\bf h}$, so its response $m$ can be represented by a tensor
${\bf R}$
such that
\begin{eqnarray}
m&=&{\bf R}\text{:}{\bf h}.
\end{eqnarray}
For our interferometric detector
\begin{eqnarray}
{\bf R}&\equiv&{1\over2}\left(
{\bf l}\otimes{\bf l}-
{\bf m}\otimes{\bf m}
\right).
\end{eqnarray}
\item Assume that axis ${\bf e}^R_x$ makes an angle $\zeta$ with
the axis ${\bf l}$, {\em i.e.,\/}
\begin{eqnarray}
{\bf e}^R_x &=& {\bf l}\cos\zeta + {\bf m}\sin\zeta.
\end{eqnarray}
\end{enumerate}
\widetext
With these conventions, the antenna patterns are given
by\cite{thorne87}
\begin{mathletters}
\begin{eqnarray}
F_{{}+{}} &\equiv& {\bf R}\text{:}{\bf e}_{{}+{}}\nonumber\\
&=&{1\over2}\cos2\zeta
\left(1+\cos^2\theta\right)\cos2\phi-
\sin2\zeta\cos\theta\sin2\phi
\label{defn:fplus}\\
F_\times &=& {\bf R}\text{:}{\bf e}_\times\nonumber\\
&=&{1\over2}\sin2\zeta
\left(1+\cos^2\theta\right)\cos2\phi+
\cos2\zeta\cos\theta\sin2\phi
\label{defn:fcross}
\end{eqnarray}
\end{mathletters}
\narrowtext
and the amplitude ${\cal A}^2$ may be written
\begin{equation}
{\cal A}^2 = {4\over{d_L}^2}\left[
F_{{}+{}}^2\left(1+\cos^2i\right)^2
+4F_\times^2\cos^2i\right].
\label{defn:scrA}
\end{equation}
It is convenient to denote the angular dependence of ${\cal A}^2$
by $\Theta^2$:
\begin{equation}
\Theta^2 = 4\left[
F_{{}+{}}^2\left(1+\cos^2i\right)^2
+4F_\times^2\cos^2i\right].
\label{defn:Theta}
\end{equation}
The range of $\Theta^2$ is $0\leq\Theta^2\leq16$.

The SNR $\rho^2$ and covariance matrix ${\cal C}_{ij}$
(eqns.~\ref{defn:rho2} and \ref{defn:cij}) both depend on ${\cal
A}^2$, which is in turn a function of the (unknown) relative
orientation of the source and interferometer through $\Theta^2$.
In order to evaluate the expected $\rho^2$ (or the expected
${\cal C}_{ij}$) of a source at a given distance ${d_L}$ we need
to know some properties of the probability distribution of
$\Theta^2$.

Since $\Theta^2$ depends on the angles $\theta$, $\phi$, $i$, and
$\zeta$ ({\em cf.\/} eqn.~\ref{defn:Theta}), the {\em a priori\/}
distribution of $\Theta^2$ depends on the {\em a priori\/}
distribution of these angles. These distributions are all known:
in particular, $\cos\theta$ and $\cos i$ are uniformly
distributed over the range $[-1,1]$ and $\phi$ and $\zeta$ are
uniformly distributed over the range $[0,2\pi)$.  Making use of
the definitions of $F_{{}+{}}$ and $F_\times$
(eqns.~\ref{defn:fplus} and \ref{defn:fcross}), we find that the
mean square of $\Theta$ is
\begin{equation}
\overline{\Theta^2} = {64\over25}.
\label{eqn:ms-Theta}
\end{equation}
The distribution of $\Theta^2$ is not symmetric, however: in
fact, its mode is zero and larger values of $\Theta^2$ are much
less likely to occur than smaller ones.  We have determined the
cumulative distribution function of $\Theta^2$ using a Monte
Carlo analysis; we give the percentiles of the distribution in
table \ref{tbl:percentiles}.  In performing these calculations,
we used Knuth's portable random number generator\cite{knuth81} as
implemented by Press {\em et al.\/}\cite{press89} ({\em i.e.,\/}
their RAN3). The results in table~\ref{tbl:percentiles} are based
on a sample of $10^7$ points in the $\{\cos\theta$, $\cos i$,
$\phi$, $\zeta\}$ parameter space. The corresponding values of
$\Theta^2$ were sorted into bins and the reported percentiles are
the rounded bin centers.

Note from table~\ref{tbl:percentiles} that significantly more
than half ({\em i.e.,\/} approximately 65\%) of the inspiralling
binary systems will have $\Theta^2$ less than
$\overline{\Theta^2}$. The skew of the distribution toward
smaller $\Theta^2$ plays a significant role when we estimate the
range of the interferometer ({\em cf.\/} \S\ref{subsec:range}).

\section{Application to LIGO}
\label{sec:ligo}

In \S\ref{sec:application} we found expressions for the SNR, the
variance, and correlation coefficients corresponding to the
detection of a signal characterized by $\widehat{{\cal A}}$,
$\widehat{{\cal M}}$, $\widehat{\psi}$, and $\widehat{T}$. In
this section we join those expressions with the design
characteristics of LIGO/VIRGO-like interferometers to obtain
estimates for the sensitivity of a realistic interferometric
detector to inspiralling binaries.

\subsection{Noise and the LIGO interferometers}
\label{subsec:noise}

The characteristic of the interferometers enter our analysis
solely through the strain noise PSD $S_h(f)$. The dominant
contributions to $S_h(f)$ for the interferometer configurations
that will be used to search for inspiralling binaries are from
seismic, thermal, and photon shot noise. In this subsection we
summarize these contributions to the overall noise PSD that we
use in our calculations.

\subsubsection{Photon shot noise}
\label{subsubsec:shot-noise}

At high frequencies $S_h(f)$ is dominated by photon shot noise.
The shot noise depends on the interferometer arm lengths, laser
power and wavelength, mirror reflectivities and configuration in
a complicated fashion. In all of our calculations we have assumed
that LIGO/VIRGO will be equipped with Fabrey-Perot cavity
interferometers, and we have used the analysis of Krolak, Lobo,
and Meers\cite[{\em cf.\/}~their eqns.~2.11-22]{krolak91} to
describe the photon shot noise.  This analysis is general enough
to encompass non-recycling, standard recycling, and dual
recycling interferometers. Except in
\S\ref{subsec:snr-discussion} we always assume that the
instrumentation makes use of standard recycling techniques. For
standard recycling, the photon shot noise is given approximately
by Thorne\cite[eqn.~117c]{thorne87} or Krolak, Lobo, and
Meers\cite[eqn.~3.5]{krolak91}
\begin{equation}
S_h^{\rm shot}(f) =
{\hbar\lambda\over\eta I_0}
{A^2\over L}f_c
\left[1+\left(f\over f_c\right)^2\right],\label{eqn:approx-shot}
\end{equation}
where $A^2$ describes the mirror losses, $I_0$ is the laser power,
$\eta$ is the quantum efficiency of the photo-detector, $\lambda$
is the laser wavelength, $L$ is the length of interferometer
arms, and $f_c$ is the recycling ``knee'' frequency.

The simplest way in which the observer can change the noise
characteristics of the LIGO/VIRGO instrumentation is by changing
the recycling knee frequency $f_c$. The general trend is that, as
the recycling frequency increases the bandwidth increases while
the sensitivity across the bandwidth
decreases\cite[fig.~9.13]{thorne87}.

\subsubsection{Thermal noise}
\label{subsubsec:thermal-noise}

At lower frequencies, off-resonance thermal excitations of the
test mass suspensions and internal modes of the pendulum masses
either dominate or provide important contributions to the noise.
We approximate the suspension noise by focusing only on the
pendulum mode (ignoring both torsional and violin modes). If the
dissipative force in the pendulum suspension is due to friction,
then the strain PSD of a single test mass $m$ with resonant
frequency $f_0$ and quality factor $Q_0$ at temperature $T$ is
given by
\begin{equation}
S^{\rm pend}_h(f) =
{k_{\rm B} T f_0\over 2\pi^3 mQ_0L^2
\left[
  \left(f^2-f_0^2\right)^2 +
  \left(f f_0/Q_0\right)^2
\right]}
\end{equation}
(note that we use $T$ for both the temperature and the Newtonian
``moment of coalescence;'' nevertheless, the meaning of the
symbol in any given context should be clear).  Each arm of the
interferometer has a pendulum degree of freedom at each end and
the noise from each degree of freedom is independent; thus, the
total noise PSD is a factor of four greater than this (note that
Dhurandhar, Krolak and Lobo\cite[eqn.~2.10]{dhurandhar89a}
neglect this factor of four in their analysis and have several
typographical errors in their formulae). The total thermal
suspension noise PSD is given by
\begin{equation}
S^{\rm susp}_h(f) = 4S_h^{\rm pend}(f).\label{eqn:sh-susp}
\end{equation}

The primary dissipative force acting on the pendulum may not be
friction, however: it has been suggested\cite{saulson90} that the
dissipation is due instead to a phase lag between the stress and
the strain in the pendulum suspension. If this is the case, then
the noise PSD is different than that given above.  This is not an
issue for the initial interferometers, but will be for the
advanced ones\cite{thorne92}. The nature of the dissipation is
far from settled, and in the absence of a consensus we have used
the form given in equation~\ref{eqn:sh-susp}.

Off-resonance thermal excitations of the vibrational modes of the
test masses will also be a significant source of noise in the
LIGO/VIRGO interferometers.  Here we consider only the
fundamental vibrational mode of each test mass.  The contribution
to the noise PSD has the same form as the thermal suspension
noise $S_h^{\rm susp}(f)$ (and is subject to the same
controversy), only now the resonant frequency and oscillator
quality that enter are those of the test masses:
\begin{equation}
S^{\rm int}_h(f) =
{2k_{\rm B} Tf_{\rm int}\over \pi^3 mQ_{\rm int}L^2
\left[
  \left(f^2 - f_{\rm int}^2\right)^2 +
  \left(f f_{\rm int}/Q_{\rm int}\right)^2
\right]}.
\end{equation}

\subsubsection{Seismic noise}
\label{subsubsec:seismic}

Seismic noise will dominate $S_h(f)$ at low frequencies.
Saulson\cite{saulson84} has surveyed the literature on the
seismic displacement noise PSD $S_x(f)$ and finds that it is
roughly proportional to $f^{-4}$ in the range
$1/10\,\text{Hz}\lesssim f\lesssim 10\,\text{Hz}$. Consequently, if the
LIGO/VIRGO test mass suspensions were coupled directly to the
Earth, then
\begin{equation}
S^{\rm seismic}_h(f) =
{S'_0f^{-4}\over\left(f_0^2-f^2\right)^2
+\left(ff_0/Q_0\right)^2}
\label{eqn:seismic-noise}
\end{equation}
where $f_0$ is the pendulum mode frequency, $Q_0$ is pendulum
quality, and $S'_0$ is a proportionality constant with units of
${\rm Hz}^{7}$. Note that at frequencies above the pendulum
frequency $f_0$ the seismic strain noise is proportional to
$f^{-8}$ while below $f_0$ is is proportional to $f^{-4}$.

In the actual LIGO/VIRGO interferometers, the pendulum
suspensions will be isolated from the Earth by a mechanical
circuit that is a series of several highly damped
oscillators\cite{ligo89,bradaschia91}. Each oscillator in this
series circuit will introduce four poles in the strain noise PSD
near the pendulum frequency; consequently, the actual seismic
noise contribution will be much steeper than $f^{-8}$ at
frequencies greater than $f_0$ and much more complicated near the
resonant frequencies of the mechanical circuit. It is proposed
that seismic isolation in the initial interferometers be provided
by a five stage circuit (where the final stage of isolation is
the pendulum suspension itself)\cite[\S{V 3 b} and appendix
D]{ligo89}; consequently, for frequencies much greater than $f_0$
the seismic strain noise PSD is expected to be proportional to
$f^{-24}$.  Nearer the resonant frequencies of the isolation
circuit the dependence is quite complicated as the poles in the
response function are not at zero frequency, but at complex
frequencies with real parts near 1~Hz. The strain noise PSD below
the resonant frequencies of the isolation circuit remains
proportional to the displacement noise PSD $S_x(f)$.

In our calculations we use the following crude estimate for the
seismic strain noise PSD:
\begin{equation}
S_h^{\rm seismic}(f) = {S_0 f^{-4}\over(f^2-f_0^2)^{10}}.
\end{equation}
The proportionality constant $S_0$ has units of Hz${}^{23}$.
This estimate scales correctly with frequency above and below
$f_0$, though it fails near $f_0$. This failure is unimportant
since (except in \S\ref{subsec:cut-off}) we assume that $f_l>f_0$
(the choice of cut-off $f_l$ is discussed in
\S\ref{subsubsec:cutoff}).

\widetext
The amplitude of the noise (as reflected by the proportionality
constant $S_0$) depends on the detailed nature of the seismic
isolation circuit and the properties of the seismic displacement
noise at the interferometer site.  The LIGO design goals for the
initial and advanced interferometers are\cite{ligo89}
\begin{mathletters}
\begin{eqnarray}
S_h^{\rm seismic}(40\,{\rm Hz}) &=&
S^{\rm susp}_h(40\,{\rm Hz})+
S^{\rm int}_h(40\,{\rm Hz})\qquad\text{(initial interferometers)}\\
S_h^{\rm seismic}(10\,{\rm Hz}) &=&
S^{\rm susp}_h(10\,{\rm Hz})+ S^{\rm int}_h(10\,{\rm Hz})
\qquad\text{(advanced interferometers)}.
\end{eqnarray}
\end{mathletters}
We fix $S_0$ by these relationships.  Where we discuss the
advanced LIGO {\em detectors,\/} we use the same condition on the
seismic noise as for the advanced interferometers.

\narrowtext
\subsubsection{Quantum noise}
\label{subsubsec:quantum}

In addition to the primary noise sources discussed above, we have
also included a contribution whose origin is quantum mechanical
and rooted in the Heisenberg uncertainty principle.  When we
observe a signal of frequency $f$ in an interferometer, we are
measuring the periodic motion of the end-masses at that
frequency.  Since the motion is periodic, this is equivalent to a
simultaneous measurement of the momentum and localization of the
end masses, and the precision with which we can make this
measurement is subject to the usual quantum mechanical limits. In
our calculations, we use the form of the quantum noise given by
Thorne\cite[eqn.~121]{thorne87}:
\begin{equation}
S_h^{\rm quant}(f) = {8\hbar\over m\left(2\pi f\right)^2L^2},
\label{eqn:quantum}
\end{equation}
where $m$ is the mass of the LIGO pendulum bobs.

\subsubsection{Noise source summary}
\label{subsubsec:noise-summary}

In table~\ref{tbl:instruments} we give the instrument
characteristics we have assumed in our calculations. Two sets of
values are given, corresponding to estimates for LIGO initial and
advanced instrumentation. These estimates have been culled from
the literature\cite{vogt91,abramovici92}, the LIGO
proposal\cite{ligo89}, and personal communication with members of
the LIGO project\cite{thorne92,whitcomb92}. In terms of the noise
sources discussed in the previous subsections, the noise PSD
$S_h(f)$ we use in our calculations is
\begin{eqnarray}
S_h(f) &=&
S_h^{\rm shot}(f) +
S_h^{\rm int}(f) +
S_h^{\rm susp}(f)
\nonumber\\
&&\quad{}+
S_h^{\rm seismic}(f) +
S_h^{\rm quant}(f).
\end{eqnarray}

As a companion to table~\ref{tbl:instruments} and as a graphical
illustration of how all of the noise sources discussed above act
in concert to determine an interferometer's noise
characteristics, we show our approximation to the anticipated
$S_h(f)$ for both the initial (fig.~\ref{fig:initial}) and
advanced (fig.~\ref{fig:advanced}) instrumentation. The
contributions to $S_h(f)$ from each of the influences discussed
above are shown as dashed lines and $S_h(f)$ is shown as a solid
line. Both figures show interferometers configured to operate in
standard recycling mode.  In figure~\ref{fig:initial},
corresponding to the initial interferometers, the recycling
frequency $f_c$ is 300~Hz, while in figure~\ref{fig:advanced}
(corresponding to the advanced interferometer design) it is
100~Hz.

\subsubsection{Choosing the low frequency cut-off $f_l$}
\label{subsubsec:cutoff}

The elements of the covariance matrix ${\cal C}_{ij}$ depend on
the moments $f_{7/3}$, $\overline{f}_{17/3}$, $\overline{f}_{4}$,
$\overline{f}_{3}$, $\overline{f}_{4/3}$, and
$\overline{f}_{1/3}$.  In turn, these depend on $S_h(f)$ and the
low frequency cut-off $f_l$ (see eqns.~\ref{defn:f7/3}
and~\ref{defn:f-beta}). Our calculations of these moments have
assumed a low frequency cut-off of 10~Hz, corresponding to the
last several minutes in the inspiral of a binary neutron star
system. In table~\ref{tbl:freq-moments} we give the frequency
moments $f_{7/3}$, $\overline{f}_{17/3}$, $\overline{f}_{4}$,
$\overline{f}_{3}$, $\overline{f}_{4/3}$, and
$\overline{f}_{1/3}$ for two cases of interest.

\subsubsection{The LIGO detector}
\label{subsubsec:ligo}

The two LIGO interferometers, though separated by several
thousand miles, share nearly the same orientation in space: the
planes defined by the detector arms are nearly parallel, and the
arms themselves are nearly parallel. Consequently the network
acts like a single interferometer of greater sensitivity than
either of its components. If the noise in the two component
interferometers of the LIGO detector is uncorrelated and
described by $S_h^{(0)}(f)$, then the effective PSD of the more
sensitive single interferometer is $S_h(f)=S^{(0)}_h(f)/2$.
Consequently, the effective PSD for the LIGO detector in the
limit that the interferometers share the same orientation is also
given by figure~\ref{fig:advanced}, but with the scale reduced by
a factor of $2^{-1/2}$. In making this approximation we are
ignoring the differences in arrival time of the gravitational
radiation signal at the two interferometers.  In the following
sections when we refer to the LIGO {\em detector\/} (as opposed
to a LIGO/VIRGO-like interferometer) we are actually referring to
a single interferometer whose noise PSD is $1/2$ that of the
advanced interferometer design.

\subsection{Signal-to-noise ratio}
\label{subsec:snr-discussion}

As discussed in \S\ref{subsec:V(P)}, we decide whether or not a
signal is present in the output of the detector by comparing the
likelihood ratio $\Lambda$ to a pre-determined threshold. In this
regard, the SNR $\rho^2$ is an acceptable surrogate for
$\Lambda$; {\em i.e.,\/} we can choose a threshold $\rho_0^2$
(which may be a function of $\widehat{\bbox{\mu}}$) to
compare with $\rho^2$.  Then, if $\rho^2\geq\rho_0^2$ we assert
the presence of a signal while if $\rho^2<\rho_0^2$ we conclude
that the detector output is only noise.  The choice of threshold
is a delicate matter: on the one hand we want a high threshold to
minimize the probability that we misidentify noise as signal; on
the other hand, we want a low threshold to minimize the
probability that we misidentify a real signal as noise.  We will
consider the proper choice of the threshold $\rho^2_0$ in a later
paper; now, however, we assume only that the threshold depends
weakly on the detection strategy and $\bbox{\mu}$.

\widetext
The amplitude SNR $\rho$ may be expressed
\begin{eqnarray}
\rho &=& 8
\left({\cal M}\over1.2\,{\rm M}_\odot\right)^{5/6}
\left(\Theta\over\Theta_{50\%}\right)
F_{7/3}
\nonumber\\
&&\qquad\times\left\{
\begin{array}{ll}
\left(17.0\,{\rm Mpc}/{d_L}\right)
&\quad\text{initial interferometers}\\
\left(308.\,{\rm Mpc}/{d_L}\right)
&\quad\text{advanced interferometers}\\
\left(436.\,{\rm Mpc}/{d_L}\right)
&\quad\text{advanced LIGO detector}
\end{array}
\right.\label{eqn:snr-sensitivity}
\end{eqnarray}
where
\begin{mathletters}
\begin{eqnarray}
F_{7/3}(f_c,f_l) &\equiv&
\left[{f_{7/3}(f_c,f_l)
\over f_{7/3}(100\,{\rm Hz},10\,{\rm Hz})} \right]^{1/2} \\
&=&\left\{
\begin{array}{ll}
\left(f_{7/3}/5.331\times10^{42}\,{\rm Hz}^{-1/3}\right)^{1/2}&
\quad\text{initial interferometers}\\
\left(f_{7/3}/1.747\times10^{45}\,{\rm Hz}^{-1/3}\right)^{1/2}&
\quad\text{advanced interferometers}\\
\left(f_{7/3}/3.493\times10^{45}\,{\rm Hz}^{-1/3}\right)^{1/2}&
\quad\text{advanced LIGO detector}
\end{array}
\right.
\end{eqnarray}
\end{mathletters}
In figure \ref{fig:f7/3} we show $F_{7/3}$ for both the initial
and advanced LIGO instrumentation. Two curves are shown: one for
the initial interferometers and one for the advanced
interferometers. Each curve assumes a low-frequency cutoff $f_l$
of 10~Hz (corresponding to the last several minutes of binary
neutron star inspiral). For the advanced instrumentation the
detection strategy that maximizes $\rho$ has $f_c=100$~Hz (where
$F_{7/3}=1$), while for the initial instrumentation $f_c=300$~Hz
(where $F_{7/3}=1.3$).  These correspond to the choice of $f_c$
in figures~\ref{fig:initial} and~\ref{fig:advanced} showing the
detailed breakdown of $S_h(f)$ for the initial and advanced
interferometers.

\narrowtext
Now consider the case of resonant dual
recycling\cite{meers88,meers89,meers91}.  The photon shot noise
in an interferometer operating in a resonant dual recycling mode
is proportional to ({\em cf.\/}~\cite{krolak91} eqn.~3.7 but note
that their approximate expression has several errors; see also
\cite{meers88,meers91})
\begin{equation}
S_h^{\rm dual}(f) \propto
\left[1+\left(f-f_n\over\Delta f\right)^2\right],
\label{eqn:dual}
\end{equation}
where $\Delta f$ and $f_n$ depend on the reflectivities of
certain mirrors in the experimental apparatus.  This shot noise
PSD is large except in a narrow band about $f_n$ where it is very
small. The bandwidth $\Delta f$ and the central frequency $f_n$
of this ``notch'' can be adjusted, and the size of $S^{\rm
dual}_h$ in and out of the notch will vary depending on the $f$
and $\Delta f$. In comparison, an interferometer operating in
standard recycling mode has a nearly constant shot noise PSD for
frequencies below the knee frequency $f_c$, with the noise PSD
increasing as $f^2$ for frequencies greater that $f_c$ ({\em
cf.\/}~eqn.~\ref{eqn:approx-shot} and figs.~\ref{fig:initial} and
\ref{fig:advanced}). For $|f-f_n|\lesssim\Delta f$, dual
recycling cuts a notch in $S_h^{\rm dual}$.  Elsewhere, $S_h^{\rm
shot}$ is lower than that for dual recycling (assuming $f_0\simeq
f_c$). On the basis of numerical investigations of the
interferometer response and the waveform of inspiralling binary
systems, Krolak, Lobo, and Meers\cite{krolak91} suggested that
dual recycling with $f_n=100$~Hz and $\Delta f\simeq6\,{\rm Hz}$
is superior to standard recycling for the observation of binary
systems.  Their analysis assumed that $S_h(f)$ is infinite below
100~Hz and that only photon shot noise is important above 100~Hz:
in particular, they did not consider the noise owing to the
thermal excitations of the test mass vibrational modes ({\em
cf.\/}~\S\ref{subsubsec:thermal-noise}). When $S_h^{\rm int}(f)$
is included in the analysis, then the advantage of dual recycling
in this regime is lost: the notch is ``filled in'' by the thermal
noise almost to the level of standard recycling photon shot noise
({\em cf.\/}~their fig.~5 and our figs.~\ref{fig:initial}
or~\ref{fig:advanced}), and everywhere else the noise is much
greater than is the case for an interferometer operated in
standard recycling mode.

Dual recycling may still be useful at much higher frequencies
where the photon shot noise is a standard recycling configuration
is much larger than the thermal noise ({\em e.g.,\/} 1.5~KHz).
Observations at such high frequencies may prove useful for
detecting the actual coalescence event\cite{cutler92a}.

\section{Astrophysical Implications}
\label{sec:implications}

\subsection{Source rate}
\label{subsec:rate}

Chernoff and Finn\cite{chernoff92} have shown that the observed
differential rate $d\dot{N}/d{\cal M} d{\cal A}$ of inspiralling
binary systems depends on the cosmological model; consequently,
it can be used to determine the Hubble constant $H$, the
deceleration parameter $q$, and otherwise distinguish between
cosmologies.  Here we are interested in the total rate of
observed binary inspiral as an estimate of the sensitivity of an
interferometer, and in this subsection we estimate that rate
ignoring cosmological effects.  We refer the reader interested in
a rate calculation consistent with an expanding universe and
taking into account evolution of the binary population and
distribution of ${\cal M}$ in binaries to Chernoff and
Finn\cite{chernoff92}.

\widetext
Assume that the rate density (number per unit co-moving
cosmological volume per unit time) of inspiralling neutron-star
binary systems is a constant $\dot{{\cal N}}$ and that the
variation in neutron star masses is small so that ${\cal M}$ is
approximately equal to $1.2\,{\rm M}_\odot$ (corresponding to two
$1.4\,{\rm M}_\odot$ neutron stars). The expected total rate
$\dot{N}$ of systems whose SNR $\rho^2$ is greater than
$\rho^2_0$ is
\begin{mathletters}
\begin{eqnarray}
\dot{N} &=& \int_0^\infty dr\,4\pi \dot{{\cal N}} r^2
P\left[\rho^2(r)>\rho_0^2\right]
\nonumber\\
&=& \int_0^\infty dr\,{4\pi} \dot{{\cal N}} r^2
P\left(\Theta^2>{r^2\over r_0^2}\right)
\nonumber\\
&=& {4\pi} \dot{{\cal N}} r_0^3 \int_0^\infty dx\,x^2
P\left(\Theta^2>x^2\right) \label{defn:x2int}\\
&\equiv& {4\pi\over3}\dot{{\cal N}}{\cal R}^3,
\end{eqnarray}
\end{mathletters}
where
\begin{mathletters}
\begin{eqnarray}
P\left(\Theta^2>x^2\right) &\equiv&
\left(\begin{tabular}{l}
The probability that $\Theta^2$ is greater than $x^2$
\end{tabular}\right)\\
r_0 &\equiv&
\left(
  5{\cal M}^{5/3}f_{7/3}\over96\pi^{4/3}\rho^2_0
\right)^{1/2}\nonumber\\
&=&
\left({\cal M}\over1.2\,{\rm M}_\odot\right)^{5/6}
\left({8\over\rho_0}\right)F_{7/3}\nonumber\\
&&\quad\times
\left\{
\begin{array}{ll}
13.0\,{\rm Mpc}&\quad\text{initial interferometer}\\
236.\,{\rm Mpc}&\quad\text{advanced interferometer}\\
334.\,{\rm Mpc}&\quad\text{advanced LIGO detector,}
\end{array}
\right.\nonumber\\
&&\\
{\cal R} &\equiv& r_0
\left[3\int_0^\infty dx\,x^2
P\left(\Theta^2>x^2\right)\right]^{1/3}
\end{eqnarray}
\end{mathletters}
[recall that we have ignored cosmological effects ({\em
cf.\/}~Chernoff and Finn\cite{chernoff92}) in
eqn.~\ref{defn:x2int}]. Using the cumulative distribution
function for $\Theta^2$ ({\em cf.\/}~\S\ref{subsec:prop-Theta}
and tbl.~\ref{tbl:percentiles}), we find that
\begin{equation}
\int^\infty_0 dx\,x^2P\left(\Theta^2 > x^2\right) = 1.84;
\end{equation}
hence,
\begin{eqnarray}
{\cal R}&=&
\left({\cal M}\over1.2\,{\rm M}_\odot\right)^{5/6}
\left(8\over\rho_0\right)F_{7/3}\nonumber\\
&&\quad\times
\left\{
\begin{array}{ll}
23.0\,{\rm Mpc}&\quad\text{initial interferometer}\\
417.\,{\rm Mpc}&\quad\text{advanced interferometer}\\
589.\,{\rm Mpc}&\quad\text{advanced LIGO detector.}
\end{array}
\right.
\end{eqnarray}
Phinney\cite{phinney91} has given estimates for the number
density of sources per unit time ($\dot{{\cal N}}$) based on
observational and theoretical arguments. These estimates range
from an ultra-conservative $6\times10^{-10}\,{\rm Mpc}^{-3}{\rm
yr}^{-1}$, to a conservative $8\times10^{-8}\,{\rm Mpc}^{-3}{\rm
yr}^{-1}$, to an upper limit of $6\times10^{-5}\,{\rm
Mpc}^{-3}{\rm yr}^{-1}$. They are based on the statistics of
local populations of binary pulsars and type Ib supernovae, and
the large range reflects both the small size of the local sample,
uncertainties in our understanding in the evolution of binary
systems, and uncertainties in the selection effects at work in
determining the fraction of the local systems we have direct
knowledge of. If we take the typical threshold $\rho_0$ to be
$8$, then we find that expected rate of detections of
inspiralling binary systems is
\begin{eqnarray}
\dot{N} &\simeq&
{\dot{{\cal N}}\over8\times10^{-8}\,{\rm Mpc}^{-3}{\rm yr}^{-1}}
\left({\cal M}\over1.2{\rm M}_\odot\right)^{5/2}
\left(8\over\rho_0\right)^{3}F^3_{7/3}\nonumber\\
&&\quad\times
\left\{
\begin{array}{ll}
4.1\times10^{-3}\,{\rm yr}^{-1}&\quad\text{initial
interferometer}\\
24.\,{\rm yr}^{-1}&\quad\text{advanced interferomter}\\
69.\,{\rm yr}^{-1}&\quad\text{advanced LIGO detector.}
\end{array}
\right.
\label{eqn:source-rate}
\end{eqnarray}
As commented earlier, to maximize the rate at which binaries are
detected we need to choose $f_c$ in order to maximize $F_{7/3}$.

For a binary system consisting of a 10~M${}_\odot$ black hole and
a neutron star, ${\cal M}\simeq3$, and for a binary system
consisting of two 10~M${}_\odot$ black holes, ${\cal M}\simeq9$
(recall that we are neglecting cosmological effects).
Consequently, for these neutron-star/black-hole
(black-hole/black-hole) binaries ${\cal R}_{90\%}\simeq2$~Gpc
(5~Gpc) for the advanced LIGO detector. The situation for
determining the rate at which such systems will be detected is a
bit more complicated. Phinney argues that black-hole/black-hole
and black-hole/neutron-star binaries form at rates comparable to
the neutron-star/neutron-star merger rate; however, the fraction
which merge depends on the model dependent details that vary
greatly\cite{phinney91}, so no reliable estimate of the
coalescence rate is available for use with
eqn.~\ref{eqn:source-rate}.

\narrowtext
\subsection{Range}
\label{subsec:range}

An important measure of the sensitivity of a LIGO/VIRGO-like
interferometer is its ``range,'' {\em i.e.,\/}~the distance to
which sources can be observed. The definition of the range is
subtle. Not all inspiralling binaries within, {\em e.g.,\/}~a
distance ${\cal R}$ will be identified as such: for some,
$\Theta$ will be less than ${\cal R}/r_0$, the corresponding SNR
$\rho^2$ will be less than the threshold $\rho_0^2$, and the
signal will be dismissed as noise. Similarly, not all
inspiralling binaries outside a distance ${\cal R}$ will fail to
be identified by the interferometer: for some $\Theta$ will be
{\em greater\/} than ${\cal R}/r_0$, the SNR $\rho^2$ will be
greater than $\rho_0^2$, and the signal will be identified as
coming from a binary system.  Since the range is a slippery
concept, we define a range {\em function\/} ${\cal R}_{\gamma}$
such that a fraction $\gamma$ of the observable sources fall
within the distance ${\cal R}_{\gamma}$:
\begin{equation}
\gamma = {
\int_0^{{\cal
R}_\gamma/r_0}dx\,x^2P\left(\Theta^2>x^2\right)\over
\int_0^\infty dx\,x^2P\left(\Theta^2>x^2\right)
}.
\label{defn:range}
\end{equation}
The quantity
\begin{equation}
{
\int_0^z      dx\,x^2P\left(\Theta^2>x^2\right)
\over
\int_0^\infty dx\,x^2P\left(\Theta^2>x^2\right)
}
\end{equation}
is tabulated in the third column of table~\ref{tbl:percentiles},
and we show $\gamma$ as a function of ${\cal R}_\gamma/{\cal R}$
in figure~\ref{fig:range}. Note that
\begin{eqnarray}
{\cal R}_{90\%} &=&
\left({\cal M}\over1.2\,{\rm M}_\odot\right)^{5/6}
\left(8\over\rho_0\right)F_{7/3}\nonumber\\
&&\quad\times\left\{
\begin{array}{ll}
37.2\,{\rm Mpc}&\quad\text{initial interferometer}\\
673.\,{\rm Mpc}&\quad\text{advanced interferometer}\\
952.\,{\rm Mpc}&\quad\text{advanced LIGO detector,}
\end{array}
\right.\nonumber\\
\end{eqnarray}
{\em {\em i.e.,\/}~for the advanced LIGO detector approximately 7
sources per year will be observed whose distance is greater
than\/} $950,{\rm Mpc}$. Like the rate, the range is sensitive to
the detection strategy so that if we wish to maximize the
sensitivity of the interferometer to either we need to choose a
detection strategy that maximizes $F_{7/3}$.

\subsection{Standard deviation and correlation coefficients}
\label{subsec:cij}

First consider the measurement of ${\cal A}$.  For all ${\cal M}$
and $f_c$ relevant for both the initial and advanced
LIGO/VIRGO-like interferometers, the fractional standard
deviation $\sigma_\eta$ ({\em cf.\/}~\ref{eqn:sigma-A}) of the
waveform amplitude ${\cal A}$ is $\rho^{-1}$; consequently, {\em
the maximum fractional one-sigma uncertainty in the determination
of ${\cal A}$ is\/}
\begin{equation}
{\sigma_{\cal A}\over\widehat{\cal A}} = 0.125 {8\over\rho_0}
\end{equation}
for both the initial and advanced interferometers.  Additionally,
the correlation coefficients $r_{{\cal A} i}$ all have magnitude
less than $10^{-4}$, indicating that ${\cal A}$ is statistically
independent of ${\cal M}$, $\psi$ and $T$ ({\em i.e.,\/}~the
random errors in measurements of ${\cal A}$ owing to detector
noise are not correlated with the corresponding errors in the
measurement of ${\cal M}$, $\psi$, or $T$).

Now turn to the measurement of ${\cal M}$. Before discussing the
exact results obtained with equation~\ref{eqn:sigma-M}, we give a
heuristic derivation of the precision with which ${\cal M}$ can
be determined. Recall that the phase $\Phi$ of the gravitational
wave signal is given by
\begin{equation}
\widetilde{\Phi} =
-2\left(\widetilde{T}-t\over5\widetilde{{\cal M}}\right)^{5/8}.
\end{equation}
The observation encompasses approximately the last ten minutes in
the life of the binary system, during which time the phase
advances by
\begin{equation}
\Delta\widetilde{\Phi} = 7.4\times10^4\left(
{\Delta t\over 10\,{\rm m}}
{1.2\,{\rm M}_\odot\over\widetilde{\cal M}}
\right)^{5/8}\,{\rm rad} .
\end{equation}
The argument of the exponential in the odds ratio
(eqn.~\ref{eqn:Lambda(mu)}) is
\begin{equation}
2\left<n,m(\bbox{\mu})\right>+
2\left<m(\widetilde{\bbox{\mu}}),
m(\bbox{\mu})\right>
-\left<m(\bbox{\mu}),m(\bbox{\mu})\right>.
\end{equation}
The contribution owing to the term
$2\left<m(\widetilde{\bbox{\mu}}),
m(\bbox{\mu})\right>$ is much greater than that owing
to the noise; consequently, to a good approximation the odds
ratio will be maximized where this quantity is maximized.
Ignoring the frequency dependence of $S_h(f)$ the term
$\left<m(\widetilde{\bbox{\mu}}),
m(\bbox{\mu})\right>$ is the correlation between two
sinusoidal functions of the phase, and is large only as long as
the advance in phase of $m(\bbox{\mu})$ is within
approximately $\pi$ radians of the advance in phase of
$m(\widetilde{\bbox{\mu}})$ over the course of the
observation. Since $\Delta\Phi$ depends only on ${\cal M}$, we
have $\pi\gtrsim|\Delta\widetilde{\Phi}-\Delta\Phi|$ or
\begin{eqnarray}
{\delta{\cal M}\over{\cal M}}&\lesssim&
{4\pi\over5}\left({\widetilde{{\cal M}}\over\Delta
t}\right)^{5/8}\nonumber\\
&\lesssim& 10^{-4}\left({\widetilde{{\cal M}}\over1.2\,{\rm
M}_\odot}{10\,{\rm m}\over\Delta t}\right)^{5/8},
\end{eqnarray}
where $\delta{\cal M}$ is $\widetilde{{\cal M}}-{\cal M}$.

\widetext
Return now to consider the exact results.
Equation~\ref{eqn:sigma-M} gives the fractional standard
deviation $\sigma_{\cal M}/\widehat{\cal M}$ in terms of the
frequency moments $\overline{f}_\beta$ and $\rho$:
\begin{eqnarray}
{\sigma_{\cal M}\over\widehat{\cal M}} &=&
\left({{\cal M}\over1.2\,{\rm M}_\odot}\right)^{5/3}
\left(8\over\rho\right)
\Sigma_{\cal M}\nonumber\\
&&\quad\times\left\{
{\arraycolsep=2\arraycolsep
\begin{array}{ll}
2.08\times10^{-4}&\quad\text{initial interferometer,}\\
2.20\times10^{-5}&\quad\text{advanced interferometer}
\end{array}
}
\right.
\label{eqn:sigma-M-numbers}
\end{eqnarray}
where
\begin{equation}
\Sigma_{\cal M} =
\left[
\left(\overline{f}_{1/3}-\overline{f}^2_{4/3}\over
\Delta\right)/
\left.
\left(\overline{f}_{1/3}-\overline{f}^2_{4/3}\over
\Delta\right)
\right|_{f_c=100\,{\rm Hz},f_l=10\,{\rm Hz}}
\right]^{1/2}.\label{defn:Sigma-M}
\end{equation}
Given a threshold $\rho_0$ such that $\rho\geq\rho_0$ for all
observed sources, equations~\ref{eqn:sigma-M-numbers}
and~\ref{defn:Sigma-M} give the maximum fractional standard
deviation in the measurement ${\cal M}$ for any binary system
observed with LIGO/VIRGO-like interferometers --- a phenomenal
precision. In interpreting eqn.~\ref{eqn:sigma-M-numbers}, note
that $\sigma_{\cal M}/\widehat{\cal M}$ is inversely proportional
to $\rho$, and recall that the SNR $\rho$ of a binary system
observed in an interferometer of the advanced design is
approximately 26 times greater than the SNR of the same binary
observed in a detector of the initial design ({\em
cf.\/}~eqn.~\ref{eqn:snr-sensitivity}).  The results for the
advanced LIGO detectors are identical to those for the advanced
interferometers, except that the amplitude SNR $\rho$ for a
binary observed in the advanced detector is $2^{1/2}$ greater
than that for the same binary observed in a single advanced
interferometer.

\narrowtext
In figure~\ref{fig:Sigma-M} we show $\Sigma_{\cal M}$ for both
the initial and advanced interferometers. For the advanced
interferometer the total variation of $\Sigma_{\cal M}$ is
approximately 20\% as $f_c$ ranges from~50~Hz to~1~KHz, while for
the initial interferometer the variation is approximately 15\%.
The optimum recycling frequency for measurement of ${\cal M}$ is
that which minimizes $\Sigma_M$, and we see that this is very
different than the choice which maximizes the number of binaries
observed ({\em cf.\/}~\S\ref{subsec:snr-discussion} and
fig.~\ref{fig:f7/3}): in fact, the optimal choice of $f_c$ for
the detection of binaries (in either the initial or advanced
interferometers) is close to the {\em worst\/} possible choice of
$f_c$ for the precise measurement of ${\cal M}$.

As we have pointed out, with observations in a single
gravitational wave interferometer the location of the source on
the sky cannot be determined. If, as has been suggested, some
coalescing binaries result in $\gamma$-ray
bursts\cite{paczynski88,narayan91}, then burst observations may
be used to localize the binary system in the sky.  The
identification between a gravitational wave burst from orbital
decay (which takes place before actual coalescence of the binary
components) and a $\gamma$-ray burst (which takes place at the
time of coalescence) depends on the accuracy with which we can
measure the time of arrival of the $\gamma$-ray burst and the
``moment of coalescence'' $T$: in all events, $T$ will be within
seconds of the actual moment of neutron star disruption and the
emission of the $\gamma$-ray burst. Consequently, we need to know
$T$, the rate of detected binary coalescence, and the rate of
$\gamma$-ray bursts (the latter both assumed to be Poisson
distributed in time) in order to evaluate the probability that a
correlation in time between a $\gamma$-ray burst and a
gravitational wave burst is coincidental.  The accuracy with
which we can determine $T$ is given by $\sigma_T$ ({\em
cf.\/}~\ref{eqn:sigma-T}):
\widetext
\begin{eqnarray}
\sigma_T &=&
\left(8\over\rho\right)
\Sigma_T\nonumber\\
&&\quad\times
\left\{
\begin{array}{ll}
1.54\times10^{-4}\,{\rm s}&\quad\text{initial interferometer,}\\
3.00\times10^{-4}\,{\rm s}&\quad\text{advanced interferometer,}
\end{array}
\right.
\end{eqnarray}
where
\begin{equation}
\Sigma_T =
\left[\left(\overline{f}_{17/3}-\overline{f}_{4}^2\over\Delta
\right)/
\left.\left(\overline{f}_{17/3}-\overline{f}_{4}^2\over\Delta
\right)\right|_{f_c=100\,{\rm Hz},f_l=10\,{\rm Hz}}
\right]^{1/2}.
\end{equation}
The results for the advanced LIGO detector are the same as those
for a single advanced interferometer.  The factor $\Sigma_T$
varies by approximately a factor of 2.5 over the range $50\,{\rm
Hz}<f_c<1000\,{\rm Hz}$, and is shown (for both the initial and
advanced interferometers) in figure~\ref{fig:Sigma-T}.  Again,
the optimal interferometer configuration for precision
measurements of $T$ is very different than that for detection of
inspiralling binaries.

The parameter $\psi$ depends on the orientation of the source and
the detector and the phase of the binary systems orbit at $t=0$.
For completeness, we also give the precision with which $\psi$
can be measured:
\begin{eqnarray}
\sigma_\psi &=&
\left(8\over\rho\right)
\Sigma_\psi\nonumber\\
&&\quad\times\left\{
\begin{array}{ll}
0.257\,{\rm rads}&\quad\text{initial interferometer,}\\
0.338\,{\rm rads}&\quad\text{advanced interferometer,}
\end{array}
\right.
\end{eqnarray}
where
\begin{equation}
\Sigma_\psi \equiv \left[
\left(\overline{f}_{17/3}-\overline{f}_4^2
\over\Delta\right)/
\left.\left(\overline{f}_{17/3}-\overline{f}_4^2
\over\Delta\right)
\right|_{f_c=100\,{\rm Hz}, f_l=10\,{\rm Hz}}
\right]^{1/2}.
\end{equation}
The results for an advanced detector are the same as those for a
single advanced interferometer.

\narrowtext
The correlation coefficients $r_{{\cal M}\psi}$, $r_{{\cal M}
T}$, and $r_{\psi T}$ are nearly independent of $f_c$ for both
the initial and advanced LIGO-like interferometers. As mentioned
above, the statistical error in ${\cal A}$ is essentially
uncorrelated with that in ${\cal M}$, $\psi$, or $T$ ({\em
i.e.,\/}~the correlation coefficients $r_{{\cal A} i}$ are for
all $\lesssim10^{-4}$).  Figure~\ref{fig:ccoeffs} shows the remaining
correlations coefficients $r_{{\cal M}\psi}$, $r_{{\cal M} T}$,
and $r_{\psi T}$ for the initial and advanced interferometers.

\subsection{The low frequency cut-off}
\label{subsec:cut-off}

In order to evaluate the covariance matrix we needed to compute
the six frequency moments $f_{7/3}$, $\overline{f}_{17/3}$,
$\overline{f}_{4}$, $\overline{f}_{3}$, $\overline{f}_{4/3}$, and
$\overline{f}_{1/3}$. The evaluation of all of these is
straightforward; however, the calculation of
$\overline{f}_{17/3}$ deserves special attention: at frequencies
below the pendulum frequency of the LIGO masses the seismic noise
PSD is proportional to $f^{-4}$ ({\em
cf.\/}~eqn.~\ref{eqn:seismic-noise}); consequently,
$\overline{f}_{17/3}$ diverges as $f_l$ approaches 0 ({\em
cf.\/}~eqn.~\ref{defn:f-beta}).

For any particular application we never encounter the divergence:
there is always a low frequency cut-off in the
integral~\ref{defn:f-beta} corresponding to the finite period of
the observation. Even if we had access through an interferometer
to the entire life history of a binary system, our model for its
evolution is relevant during only a small part of its lifetime:
for example, we have assumed that the orbit is circular for all
times (when in fact gravitational radiation may be responsible
for circularizing it), that the evolution of the orbit is due
exclusively to the gravitational forces acting between the two
components, and that the two bodies are bound in a binary into
the infinite past.  Similarly, our model of the detector noise
PSD $S_h(f)$ is not necessarily valid at very low frequencies.
Nevertheless, the divergence of $\overline{f}_{17/3}$ tells us
something interesting about the observation of a binary system in
gravitational radiation and it is worthwhile to spend a few
moments understanding its origin.  So, for the purpose of
understanding this divergence we assume that we know $S_h(f)$ as
$f$ tends to 0, that our model for the waveform from a binary is
correct throughout its life history, and that the lifetime of a
binary extends into the infinite past.

Turn first to the related example of a strictly monochromatic
signal of frequency $f_0$:
\begin{equation}
m(t) = A\cos2\pi f_0t
\label{eqn:mono}
\end{equation}
for all $t$.  The noise PSD of the detector is given by $S_h(f)$.
If we observe the signal for a finite period of time $\tau$, the
SNR $\rho^2$ is
\begin{equation}
\rho^2 = {\tau A^2\over S_h(f_0)}.
\end{equation}
No matter how small $A^2$ is compared to the power in the noise,
the signal can always be discerned given a long enough
observation time.  The situation with binary inspiral is similar:
consider the transformation of the time coordinate\cite{smith87}
\begin{equation}
t'\equiv -{\cal M}\left(T-t\over5{\cal M}\right)^{5/8}.
\end{equation}
In terms of $t'$, the signal can be expressed
\begin{mathletters}
\begin{eqnarray}
h(t) &=& {{\cal A}{\cal M}\over4}
\left(-{{\cal M}\over t'}\right)^{5/32}
\cos(\Phi+\varphi)\\
\Phi(t') &=& 2\pi f't'
\label{eqn:binary-stretched}\\
f' &=& {1\over2\pi{\cal M}}.
\label{eqn:freq-stretched}
\end{eqnarray}
\end{mathletters}
Thus, the signal from an inspiralling binary is very much like
that from a monochromatic source of radiation, save that
\begin{enumerate}
\item The signal amplitude tends to zero as the stretched time
$t'$ tends to $-\infty$;
\item The detector noise amplitude tends to $\infty$ as the
stretched time $t'$ tends to $-\infty$; and
\item The signal ends at $t'=0$.
\end{enumerate}
As a result, as long as the ratio of the signal amplitude to the
noise PSD does not decrease too rapidly with decreasing frequency
(the precise rate determined by the rate at which the frequency
changes with time), then $\rho^2$ for a inspiralling binary
system should increase without bound as the observation period
extends into the infinite past. This is the role that the moment
$f_{7/3}$ plays in equation~\ref{defn:rho2} for $\rho^2$: if the
detector noise PSD increases as or less rapidly than $f^{-7/3}$
as $f\to0$, then $f_{7/3}$ diverges as the observation period is
extended into the infinite past ({\em i.e.,\/}~as the cut-off
frequency $f_l$ tends to zero) and the SNR increases without
bound.  In the case of LIGO, the PSD owing to seismic noise
increases as $f^{-4}$ at frequencies below the suspension
pendulum frequency, so that even an infinite observation period
leads to a finite SNR.

Like the SNR, the frequency $f_0$ of a truly monochromatic signal
({\em i.e.,\/}~eqn.~\ref{eqn:mono}) can be determined to
arbitrary precision given a sufficiently long observation period.
By analogy, this is equivalent to the determination of the mass
parameter ${\cal M}$ of a inspiralling binary system ({\em
cf.\/}~eqn.~\ref{eqn:freq-stretched}).  Consequently, we expect
that as long as the ratio of the signal amplitude to the noise
PSD does not increase too rapidly, the variance in the ${\cal M}$
decreases to zero as the observation period extends into the
infinite past.  Too rapidly, in this case, is $f^{-17/3}$.  Thus,
{\em even though the signal power in a given bandwidth may be
much lower than the noise power in the same bandwidth, the
information present can still play an important role in
determining the precision with which the parameterization of the
signal can be determined.}

Again, we emphasize that these conclusions refer only to the
idealized case of a circular binary system of two point masses
evolving exclusively owing to the emission of quadrupole
gravitational radiation. The relevance of these conclusions is
that the limit $f_l\to0$ ({\em
i.e.,\/}~$\overline{f}_{17/3}\to\infty$), which may seem far from
the reality of observation, is in fact very close to that which
can be attained in LIGO {\em operating in a regime where all our
approximations are valid.}

In the limit $f_l\to0$ ($\overline{f}_{17/3}\to\infty$), the
variance $\sigma^2_\zeta$ and the correlation coefficients
$r_{i\zeta}$ vanish, corresponding to the determination of ${\cal
M}$ to infinite precision. The remaining variances are ({\em
cf.\/}~eqns.~\ref{eqn:sigma-A},
\ref{eqn:sigma-psi}, \ref{eqn:sigma-T})
\begin{mathletters}
\begin{eqnarray}
{}_\infty\sigma^2_{\eta} &=& \rho^{-2}\\
{}_\infty\sigma^2_\psi &=&
{\overline{f}_{1/3}\over
\overline{f}_{1/3}-\overline{f}_{4/3}^2}{1\over\rho^2}\\
{}_\infty\sigma^2_T &=&
\left[
  4\pi^2\rho^2\left(
    \overline{f}_{1/3}-\overline{f}_{4/3}^2
  \right)
\right]^{-1},
\end{eqnarray}
\end{mathletters}
and the remaining correlation coefficients are ({\em
cf.\/}~eqns.~\ref{eqn:r-A-psi}, \ref{eqn:r-A-T},
\ref{eqn:r-psi-T})
\begin{mathletters}
\begin{eqnarray}
{}_\infty r_{\eta\psi} &=& 0\\
{}_\infty r_{\eta T} &=& 0\\
{}_\infty r_{\psi T} &=&
{\overline{f}_{4/3}\over\overline{f}^{1/2}_{1/3}}.
\end{eqnarray}
\end{mathletters}
In this limit, the moments $f_{7/3}$, $\overline{f}_{4/3}$, and
$\overline{f}_{1/3}$ describe completely the precision with which
${\cal A}$, $\psi$, and $T$ can be measured. For LIGO/VIRGO-like
interferometers, these moments change negligibly when we pass
from $f_l=10$~Hz to the limit of $f_l=0$~Hz [recall that we are
assuming $S_h(f)\propto f^{-4}$ at frequencies below the resonant
frequencies of the seismic isolation circuit]; consequently, in
observing more than the last several minutes of binary inspiral
the SNR of the observed signal is unchanged, and the variances
and correlation coefficients are independent of the details of
the ultralow frequency behavior of the interferometers.

\widetext
By analogy with $\Sigma_\psi$ and $\Sigma_T$ we define
${}_\infty\Sigma_\psi$ and ${}_\infty\Sigma_T$:
\begin{mathletters}
\begin{eqnarray}
{}_\infty\sigma_\psi &=&
\left(8\over\rho\right)
{}_\infty\Sigma_\psi\left\{
\begin{array}{ll}
0.232\,{\rm rads}&\quad\text{initial interferometer,}\\
0.190\,{\rm rads}&\quad\text{advanced interferometer,}
\end{array}
\right.\\
{}_\infty\sigma_T &=&
\left(8\over\rho\right)
{}_\infty\Sigma_T\left\{
\begin{array}{ll}
1.32\times10^{-4}\,{\rm s}&\quad\text{initial interferometer}\\
2.71\times10^{-4}\,{\rm s}&\quad\text{advanced interferometer,}
\end{array}
\right.\\
{}_\infty\Sigma_\psi &=&
\left[
\left(
\overline{f}_{1/3}\over
\left(\overline{f}_{1/3}-\overline{f}_{4/3}^2\right)
\right)/
\left.\left(
\overline{f}_{1/3}\over
\left(\overline{f}_{1/3}-\overline{f}_{4/3}^2\right)
\right)\right|_{f_c=100\,{\rm Hz},\,f_l=10\,{\rm Hz}}
\right]^{1/2}\\
{}_\infty\Sigma_T &=&
\left\{
\left.
\left[\left(\overline{f}_{1/3}-\overline{f}_{4/3}^2\right)\right]
\right|_{f_c=100\,{\rm Hz},\,f_l=10\,{\rm Hz}}
\over
\left(\overline{f}_{1/3}-\overline{f}_{4/3}^2\right)
\right\}^{1/2}.
\end{eqnarray}
\end{mathletters}
The factor ${}_\infty\Sigma_T$ is shown together with $\Sigma_T$
in figure~\ref{fig:Sigma-T} for both the initial and advanced
interferometers. Over the range $50\,{\rm Hz}<f_c<1\,{\rm Khz}$
the differences between ${}_\infty\Sigma_i$ and $\Sigma_i$ are
small.

\narrowtext
Since the SNR is unchanged as the observation period expands from
the last several minutes of binary inspiral to include the entire
lifetime of the binary, observations over more than the last
several minutes of the lifetime of a binary system will have an
insignificant effect on the number of binaries observed ({\em
cf.\/}~eqn.~\ref{eqn:source-rate}).  Comparing the expressions
given above for ${}_\infty\sigma_i$ with those for $\sigma_i$
({\em cf.\/}~eqns.~\ref{eqn:sigma-A}--\ref{eqn:sigma-T}) shows
that increasing the observation period increases the precision
with which $T$ can be determined by approximately 10\% and the
precision with which $\psi$ can be measured a factor of two.

\section{Conclusions}
\label{sec:conclusions}

Inspiralling binary systems of compact objects are regarded as
the most certain observable source of gravitational radiation for
the the Laser Interferometer Gravitational-wave Observatory
(LIGO). As a start toward understanding the capabilities of the
LIGO instruments in observations of inspiralling binary systems,
we have investigated the sensitivity of a single interferometer
of the LIGO type to the gravitational radiation from inspiralling
binary systems in the quadrupole approximation.

Observation of binary inspiral in a single LIGO/VIRGO-like
interferometer can, in principle, determine a characteristic mass
${\cal M}$, signal amplitude ${\cal A}$, time $T$, and phase
$\psi$.  The mass ${\cal M}$ is a function only of the masses of
the system's components and its cosmological redshift. The
amplitude ${\cal A}$ is inversely proportional to its luminosity
distance and depends also on a function of four angles describing
the relative orientation of the binary and the interferometer.
The time $T$ is related to the moment of binary coalescence.
Finally, $\psi$ is related to the phase of the binary system at a
fixed moment of time and is also a function of the relative
orientation angles.

The probability that the detector response is consistent with the
presence of a signal from an inspiralling binary system is
related to the signal-to-noise ratio (SNR) $\rho$ that
characterizes the observation. In practice, a threshold $\rho_0$
is chosen and we assert that a signal is present in the detector
output only if $\rho\geq\rho_0$. We characterize our uncertainty
in the parameters $\widetilde{\bbox{\mu}} =
\{\widetilde{{\cal A}}$, $\widetilde{{\cal M}}$, $\widetilde{\psi}$,
$\widetilde{T}\}$ that describe the detected binary system by
defining volumes $V(P)$ in parameter space such that
$\widetilde{\bbox{\mu}}\in V(P)$ with probability $P$.

When $\rho^2$ is large, the probability density from which $V(P)$
is constructed is a multivariate Gaussian. Consequently, the
determination of $V(P)$ is equivalent to the determination of the
several variance and correlation coefficients that describe the
Gaussian. These coefficients in turn describe the statistical
uncertainty in the determination of ${\cal A}$, ${\cal M}$,
$\psi$, and $T$, and the correlation in the errors in each.

For observations of binary systems in LIGO/VIRGO-like
interferometers, the expected SNR, variance, and correlations
coefficients may be expressed in terms of the mode of the
probability distribution $P(\bbox{\mu})$ and several
moments of the noise PSD of the interferometer We have used a
detailed model of the PSD for both the initial and advanced LIGO
interferometers configured for standard recycling, and have
evaluated the moments of the PSD, the expected SNR, variances,
and correlation coefficients as functions of the recycling knee
frequency.

The two interferometers of the LIGO detector share nearly the
same orientation. Consequently they will act similarly to a
single, more sensitive interferometer. In addition to providing
results for a single interferometer of the LIGO/VIRGO type
(either initial or advanced), we also express our results for the
LIGO two-interferometer network in the limit that the
interferometers share exactly the same orientation.

{}From the expected SNR and an estimate for the cosmological rate
density of inspiralling binary systems we have calculated the
rate of observed binary inspiral events as a function of the SNR
threshold. {\em We find that for the advanced LIGO detector a
conservative estimate of the rate of observed binary neutron star
inspiral events is 69~${\rm yr}^{-1}$, of which 7 per year will
be from binaries at distances greater than 950~Mpc.\/} This is
important for observational cosmology, since the differential
rate ({\em i.e.,\/} $d\dot{N}/d{\cal A} d{\cal M}$) depends on
the Hubble constant and other cosmological
parameters\cite{chernoff92}.

For observed binary systems, the fractional standard deviation in
the characteristic waveform amplitude is equal to $1/\rho$: if
the threshold $\rho$ is 8, then the fractional one-sigma
uncertainty in the measured amplitude will be less than 12.5\%
for sources observed in either LIGO or LIGO-like interferometers.
The chirp mass can be measured to phenomenal precision: again, if
the threshold $\rho$ is 8 then the fractional one-sigma
uncertainty in ${\cal M}$ will be less than $2.2\times10^{-5}$
for binary neutron star systems observed in the advanced LIGO
detector.  We have also calculated the precision with which $T$
and $\psi$ can be determined.

The optimum detector configuration for the observation of binary
inspiral depends sensitively on the goal of the observation.  For
example, if the object is to maximize the rate of observed binary
systems without constraining the uncertainties in ${\cal A}$,
${\cal M}$, $T$ and $\psi$, then one detector configuration is
clearly favored. On the other hand, if the object is to be able
to characterize as precisely as possible one of the observables
({\em e.g.,\/} ${\cal M}$), while allowing that some otherwise
observable sources may be missed entirely, then another detector
configuration is preferred. We have given a concrete formulation
to the question of optimum interferometer configuration and
answered it in the context of our model for the interferometer
noise PSD.

The quadrupole approximation is useful for our LIGO/VIRGO
appraisals; however, the neglect of post-Newtonian contributions
(including spin-orbit and spin-spin interactions) to the
gravitational radiation luminosity and waveform is a weakness of
our estimates and should be remedied in a more detailed appraisal
of interferometer sensitivity\footnote{Preliminary Monte Carlo
investigations by Cutler\cite{cutler92c} suggest that the
inclusion of some of the terms neglected in our analysis increase
the fractional one-sigma uncertainty in ${\cal M}$ by no more
than a factor of ten over our estimate, and have a much smaller
effect on ${\cal A}$, $T$, and $\psi$.}. Including these
interactions will increase the information that can be extracted
from gravitational radiation observations of binaries over that
described here.  The formalism that we have developed --- where
the SNR, variances, and correlation coefficients are all
expressed in terms of moments of the interferometer noise PSD ---
should prove valuable in that regard: it is readily extended to
encompass an arbitrarily more sophisticated gravitational
radiation waveform that is richer in information regarding the
source than the one we have studied here.

\acknowledgments

We would like to thank Curt Cutler, Eanna Flanagan, Tom Loredo,
Kip Thorne, Ira Wasserman, and Stan Whitcomb for helpful
discussions. L.~S.~Finn would like to thank the Alfred P.~Sloan
Foundation for their generous support, and D.~Chernoff would like
to thank the National Science Foundation for his support as a
Presidential Young Investigator. This work was supported at
Northwestern University by NASA grant NAGW-2936 and at Cornell
University by NSF grant AST-8657467 and NASA grant NAGW-2224.



\begin{figure}
\caption{The noise power spectral density (PSD) $S_h(f)$ for the
(anticipated) initial LIGO interferometers configured for
standard recycling with a knee frequency of 300~Hz. The solid
line shows the total PSD, while the dashed lines show the
important physical limits and environmental influences that
contribute to the total. For more detail, see
Sec.~\protect{\ref{subsec:noise}} and table
\protect{\ref{tbl:instruments}}.}
\label{fig:initial}
\end{figure}

\begin{figure}
\caption{The noise power spectral density (PSD) $S_h(f)$ for the
(anticipated) advanced LIGO interferometers configured for
standard recycling with a knee frequency of 100~Hz. The solid
line shows the total PSD, while the dashed lines show the
important physical limits and environmental influences that
contribute to the total. In the limit that the two LIGO
interferometers have identical orientations and we ignore the
information available owing to gravitational wave burst arrival
time differences, the (advanced design) LIGO detector noise PSD
$S_h(f)$ is $1/2$ the value shown here.  For more detail, see
Sec.~\protect{\ref{subsec:noise}} and table
\protect{\ref{tbl:instruments}}.}
\label{fig:advanced}
\end{figure}

\begin{figure}
\caption{The sensitivity of a LIGO-like interferometer to the
gravitational radiation from a coalescing binary system depends
on the detailed characteristics of the interferometer through
several moments of the inverse of the its power spectral density
(PSD) $S_h(f)$. In particular, the signal-to-noise ratio (SNR)
$\rho^2$ depends on a moment of $S_h^{-1}(f)$. Here we show how
this moment (normalized to its value at a recycling frequency of
100~Hz) varies with the choice of interferometer recycling
frequency. To maximize the rate at which sources are detected
this quantity should be maximized. For more details see
\protect{\ref{subsec:snr-discussion}}.}
\label{fig:f7/3}
\end{figure}

\begin{figure}
\caption{We define the range function ${\cal R}_\gamma$ of a LIGO-like
interferometer as the distance ${d_L}$ within which a fraction
$\gamma$ of the observable sources are expected to lie.  We also
define a characteristic distance ${\cal R}$, such that the total
rate of observable sources is $4\pi{\cal R}^3\dot{{\cal N}}/3$,
where $\dot{{\cal N}}$ is the rate density of sources (which we
assume to be uniform). A conservative estimate of ${\cal R}$ for
an advanced LIGO-like interferometer is 420~Mpc. Here we show
$\gamma$ as a function of ${\cal R}_\gamma/{\cal R}$. For further
discussion see
Sec.~\protect{\ref{subsec:range}}.}
\label{fig:range}
\end{figure}

\begin{figure}
\caption{The fractional standard deviation $\sigma_{\cal
M}/\widehat{\cal M}$ of the measured mass ${\cal M}$ depends on
the distance to the source, the relative orientation of the
source and the interferometer, and a factor $\Sigma_{\cal
M}(f_c)$ that depends on the interferometer configuration ({\em
i.e.,\/}~the recycling frequency $f_c$; {\em
cf.\/}~eqn.~\protect{\ref{defn:Sigma-M}} and
Sec.~\protect{\ref{subsec:cij}}).  Here we show $\Sigma_{\cal M}$
as a function of $f_c$ for initial and advanced LIGO-like
interferometers. In order to maximize the precision with which
${\cal M}$ can be determined, the recycling frequency should be
chosen to minimize $\Sigma_{{\cal M}}$. The corresponding
recycling frequency differs from that which should be chosen to
maximize the {\em rate\/} of sources detected ({\em
cf.\/}~fig.~\protect{\ref{fig:f7/3}}).  For more details, see
Sec.~\protect{\ref{subsec:cij}}.}
\label{fig:Sigma-M}
\end{figure}

\begin{figure}
\caption{The characteristic time $T$ (related to the moment when
coalescence occurs) can be determined from pre-coalescence
observations. The precision with which it can be determined
depends on (among other things) a factor $\Sigma_T$ (defined in
Sec.~\protect{\ref{subsec:cij}}) that encapsulates all dependence of the
precision on the duration of the observation and the
characteristics of the detector noise. The duration of the
observation will typically be several minutes, during which time
the radiation frequency will sweep from 10~Hz to approximately
1~Khz, and $\Sigma_T$ assumes that the observation period is
limited in this way. Here we show $\Sigma_T$ as a function of the
detector recycling frequency for both initial and advanced
LIGO-like interferometers. An interferometer optimized to measure
$T$ as precisely as possible should minimize $\Sigma_T$.  Also
shown are factors ${}_\infty\Sigma_T$ for initial and advanced
interferometers. These are defined identically to $\Sigma_T$
except that they assume an observation period that extends over
the entire life of an idealized binary system evolving only owing
to the emission of gravitational radiation.  The small
differences between ${}_\infty\Sigma_T$ and $\Sigma_T$ show that
the characteristics of the interferometer do not change
significantly as the observation period is lengthened beyond the
last several minutes of binary inspiral. For more details, see
Sec.~\protect{\ref{subsec:cut-off}}.}
\label{fig:Sigma-T}
\end{figure}

\begin{figure}
\caption{The correlation coefficients describe correlations among the
statistical errors in the measurements of the parameters
describing an inspiralling binary system. There is no correlation
between errors in the measurement of the waveform characteristic
amplitude ${\cal A}$ and any of the other measurable parameters.
The remaining correlation coefficients are shown here, for both
initial and advanced LIGO-like interferometers, as a function of
the interferometer recycling frequency. For more details, see
Sec.~\protect{\ref{subsec:cij}}.}
\label{fig:ccoeffs}
\end{figure}

\narrowtext
\begin{table}
\caption{The amplitude of the gravitational radiation waveform
observed in a single detector depends on the relative orientation
between the source and the detector through the function
$\Theta^2$ ({\em cf.}~eqn.~\protect{\ref{defn:Theta}}). The
orientation angles are unknown and cannot be acertained by
observation; however, their {\em a priori\/} distribution is
known and consequently the {\em a priori\/} distribution of the
signal amplitude for binaries at a fixed distance from the
detector is also known. In this table we give the cumulative
probability distribution of $\Theta^2$, and also a function of
$P(\Theta^2>x^2)$ that arises when we evaluate the number of
sources within a given distance whose signal amplitudes exceed a
given threshold. For more details, see
Sec.~\protect{\ref{subsec:prop-Theta}},
\protect{\ref{subsec:rate}}, and \protect{\ref{subsec:range}}.}
\label{tbl:percentiles}
\begin{tabular}{ddd}
$P(\Theta^2>x^2)$&$x^2$&
$\int_0^x dz\,z^2P(\Theta^2>z^2)\over
\int_0^\infty dz\,z^2P(\Theta^2>z^2)$\\
\tableline\\
90.0\%&0.240&2.00\%\\
80.0\%&0.542&6.32\%\\
75.0\%&0.707&9.06\%\\
70.0\%&0.878&12.06\%\\
60.0\%&1.250&18.82\%\\
50.0\%&1.709&27.11\%\\
40.0\%&2.283&36.98\%\\
30.0\%&3.020&48.29\%\\
25.0\%&3.485&54.54\%\\
20.0\%&4.063&61.37\%\\
10.0\%&6.144&79.48\%\\
9.0\%&6.471&81.59\%\\
8.0\%&6.832&83.75\%\\
7.0\%&7.239&85.94\%\\
6.0\%&7.701&88.18\%\\
5.0\%&8.233&90.42\%\\
4.0\%&8.857&92.64\%\\
3.0\%&9.614&94.82\%\\
2.0\%&10.589&96.91\%\\
1.0\%&11.985&98.77\%\\
0.5\%&13.054&99.52\%\\
0.4\%&13.350&99.65\%\\
0.3\%&13.682&99.77\%\\
0.2\%&14.091&99.87\%\\
0.1\%&14.284&99.95\%\\
\end{tabular}
\end{table}
\mediumtext
\begin{table}
\caption{The sensitivity of LIGO depends in large measure on the
details of the instrumentation, and this is expected to evolve
over the lifetime of the physical plant. We consider the two
extremes of detector technology that have been cited by the LIGO
team in their reports\protect{\cite{vogt91,abramovici92}}: the
instrumentation that is expected to be available when the
facilities first come online (initial interferometers), and that
expected to be available much later (advanced interferometers).}
\label{tbl:instruments}
\begin{tabular}{lll}
{}&Initial&Advanced\\
{}&interferometers&interferometers\\
\tableline\\
Temperature (T)&$300^\circ{\rm K}$&$300^\circ{\rm K}$\\
Pendulum frequency ($f_0$)&1 Hz&1 Hz\\
Suspension Quality ($Q_0$)&$10^7$&$10^9$\\
End mass ($m$)&10 Kg&1000 Kg\\
End mass fundamental mode ($f_{\rm int}$)&23 KHz&5 KHz\\
End mass quality ($Q_{\rm int}$)&$10^5$&$10^5$\\
Effective laser power ($I_0\eta$)&5 W&60 W\\
Laser wavelength ($\lambda$)&5139\AA&5139\AA\\
Mirror losses ($A^2$)&$5\times10^{-5}$&$2\times10^{-5}$\\
\end{tabular}
\end{table}
\squeezetable
\begin{table}
\caption{The signal-to-noise ratio, the variances, and the
correlation coefficients all depend on the characteristics of the
interferometer through the moments of its inverse noise power
spectral density ({\em cf.\/} Sec.~\protect{\ref{subsec:noise}}).
Here we give
those moments for the initial and advanced LIGO instrumentation
for the two cases case where the standard recycling ``knee''
frequency $f_c$ is 100~Hz and 500~Hz. In both cases, the
low-frequency cut-off (determined by the duration of the
observation) is 10~Hz.}\label{tbl:freq-moments}
\begin{tabular}{dddddd}
&&\multicolumn{2}{c}{Initial}&\multicolumn{2}{c}{Advanced}\\
&&\multicolumn{2}{c}{Interferometers}&
\multicolumn{2}{c}{Interferometers}\\
$f_c$&$[Hz]$&
\multicolumn{1}{c}{100}&\multicolumn{1}{c}{500}&
\multicolumn{1}{c}{100}&\multicolumn{1}{c}{500}\\
\tableline\\
$f_{7/3}$&$[{\rm Hz}^{-1/3}]$&
5.331$\times10^{42}$&6.520$\times10^{42}$&
1.747$\times10^{45}$&1.353$\times10^{45}$\\
$\overline{f}_{17/3}$&$[{\rm Hz}^{-10/3}]$&
9.540$\times10^{-8}$&7.704$\times10^{-8}$&
6.758$\times10^{-6}$&8.267$\times10^{-6}$\\
$\overline{f}_{4}$&$[{\rm Hz}^{-5/3}]$&
2.059$\times10^{-4}$&1.721$\times10^{-4}$&
1.562$\times10^{-3}$&1.720$\times10^{-3}$\\
$\overline{f}_{3}$&$[{\rm Hz}^{-2/3}]$&
3.056$\times10^{-2}$&2.775$\times10^{-2}$&
6.615$\times10^{-2}$&6.741$\times10^{-2}$\\
$\overline{f}_{4/3}$&$[{\rm Hz}]$&
2.346$\times10^{2}$&2.939$\times10^{2}$&
8.423$\times10$&9.536 $\times10$\\
$\overline{f}_{1/3}$&$[{\rm Hz}^{2}]$&
7.765$\times10^{4}$&1.394$\times10^{5}$&
1.248$\times10^{4}$&2.282$\times10^{4}$
\end{tabular}
\end{table}

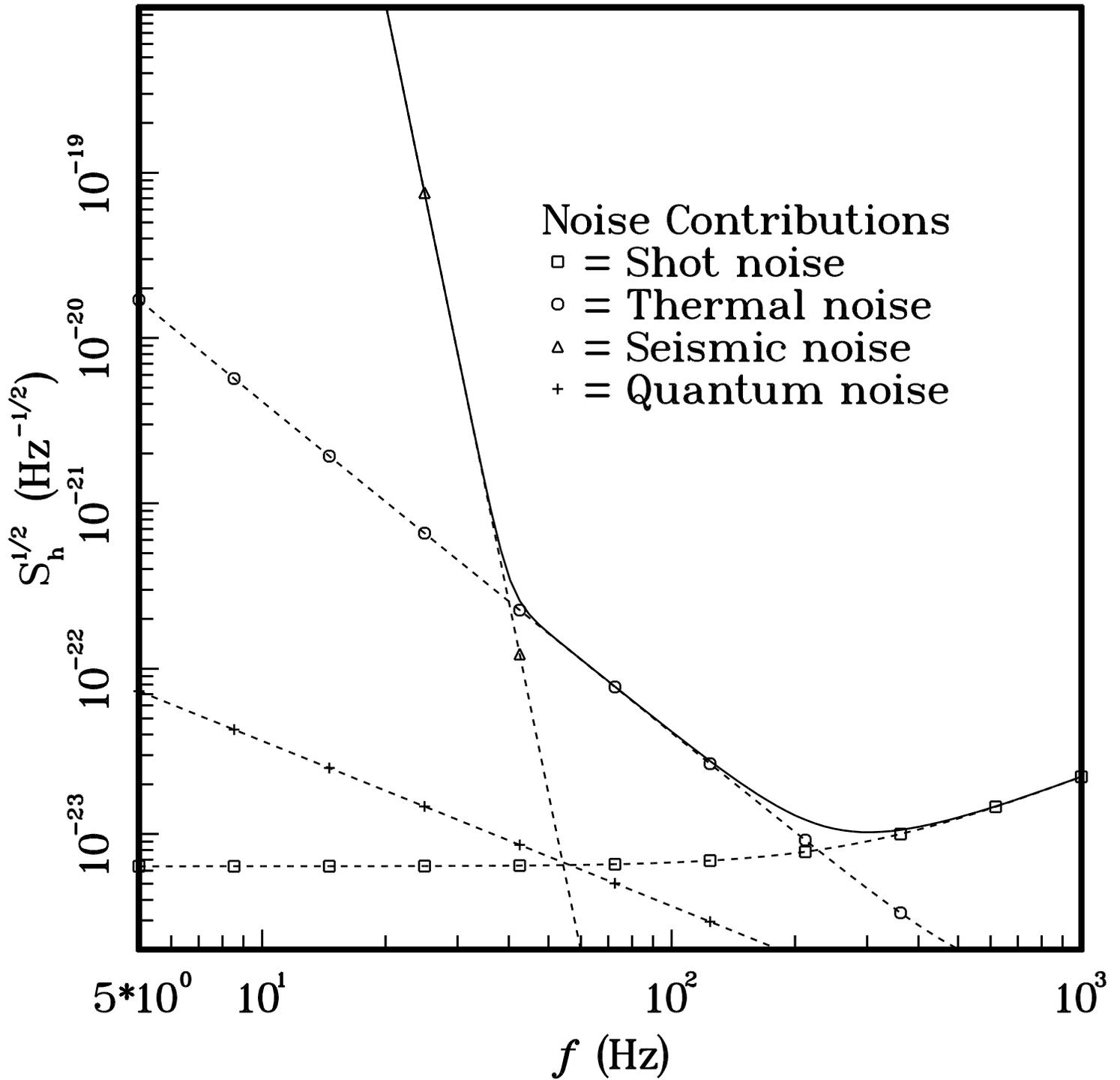
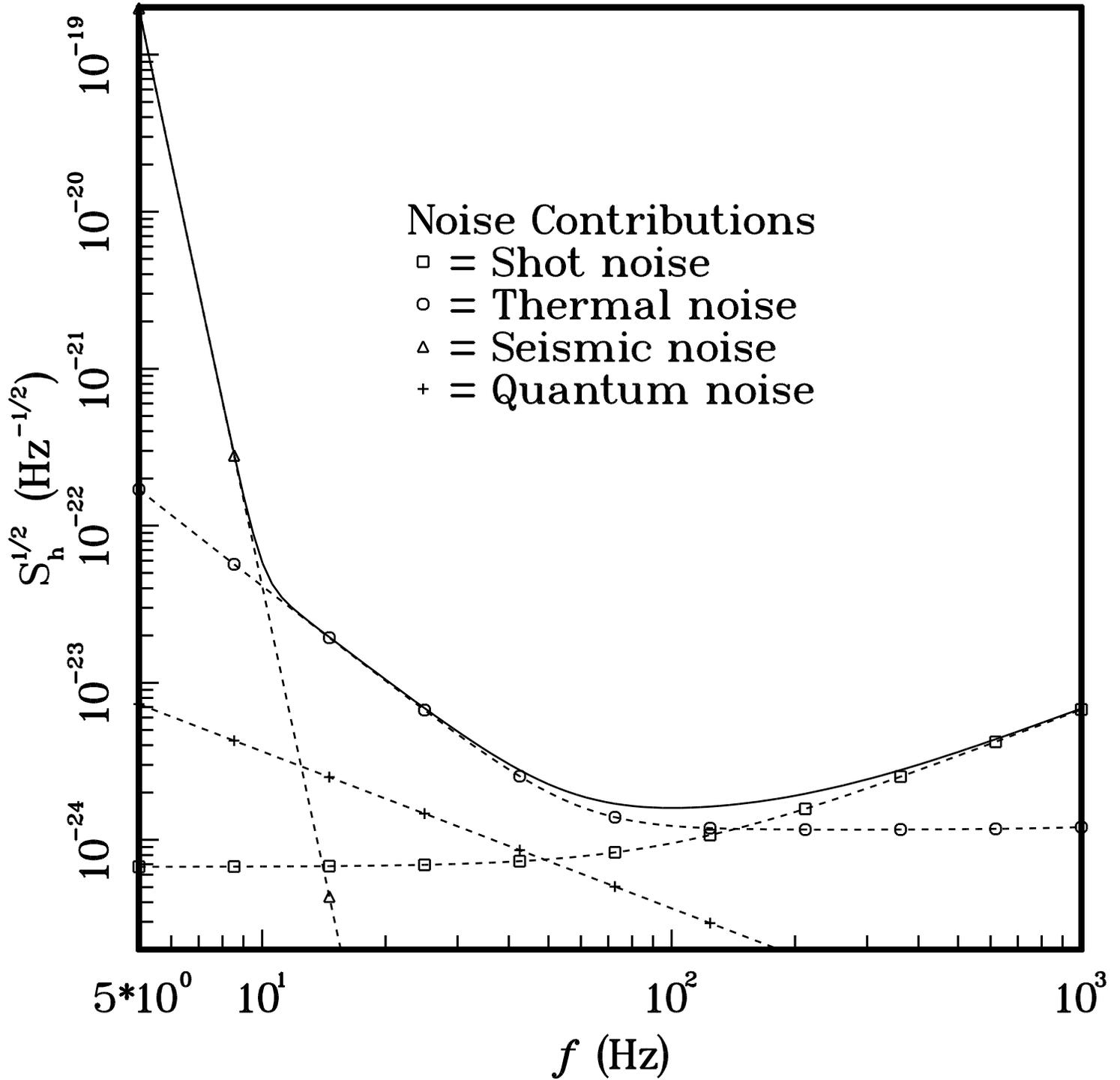
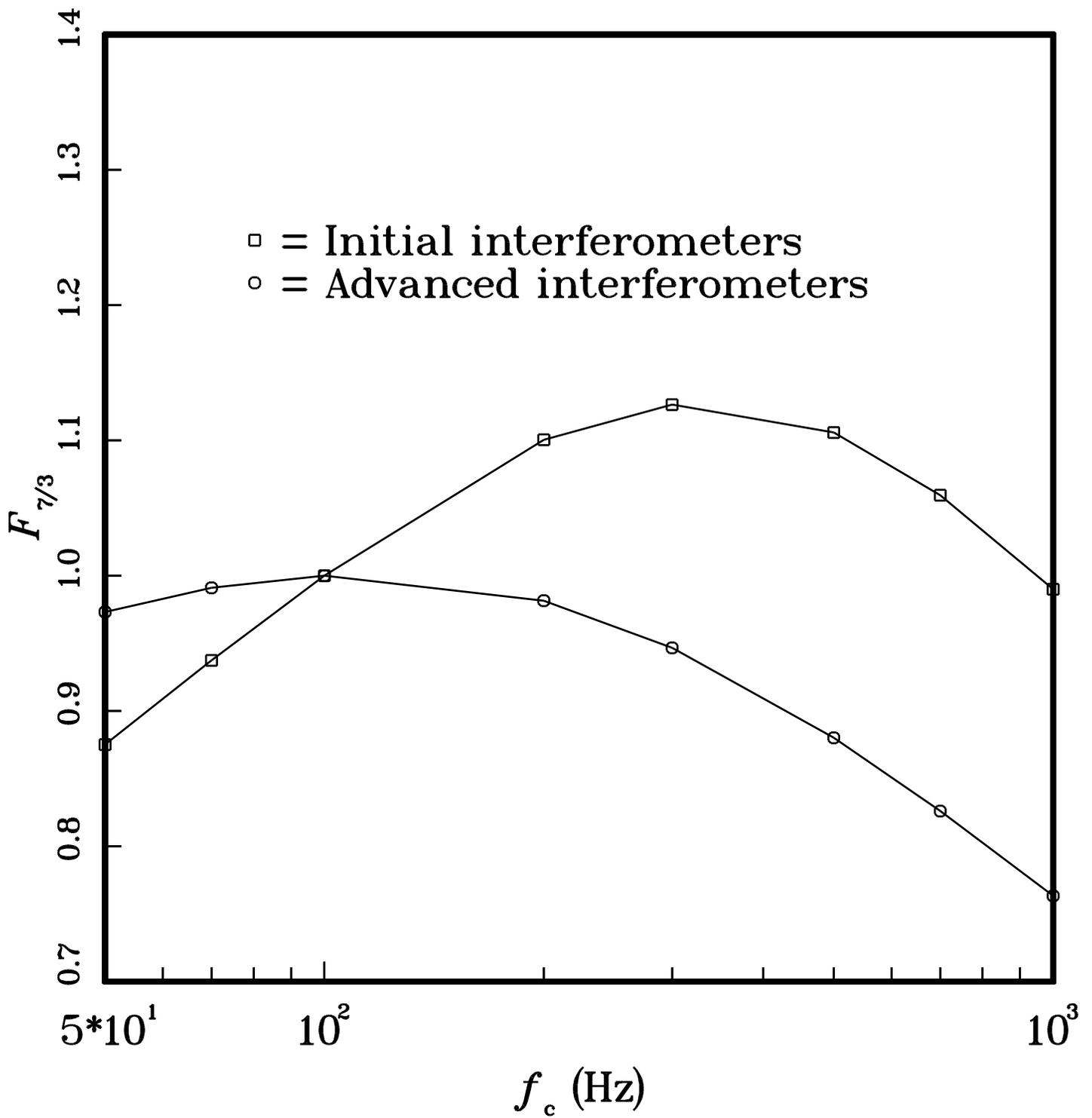
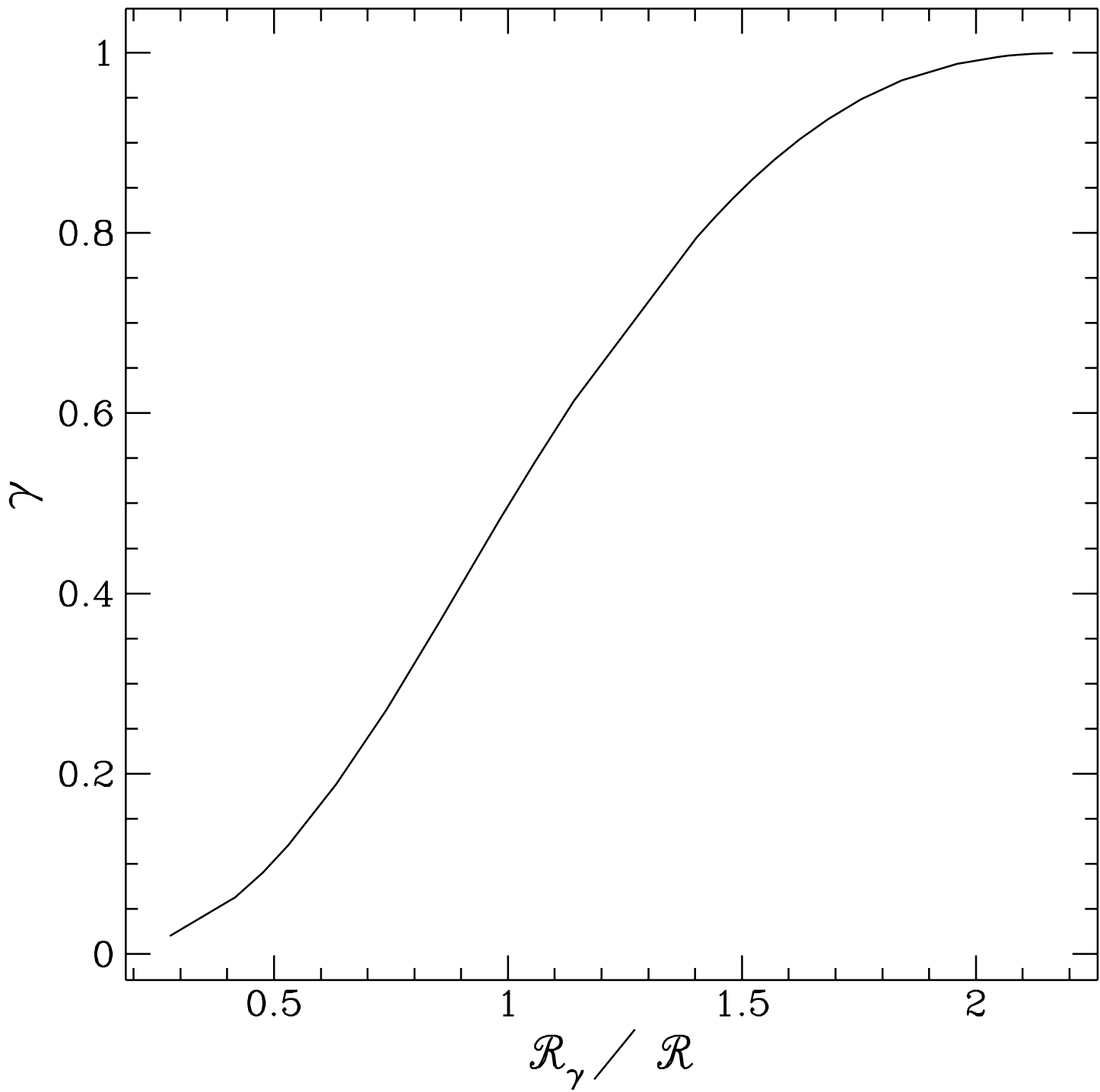
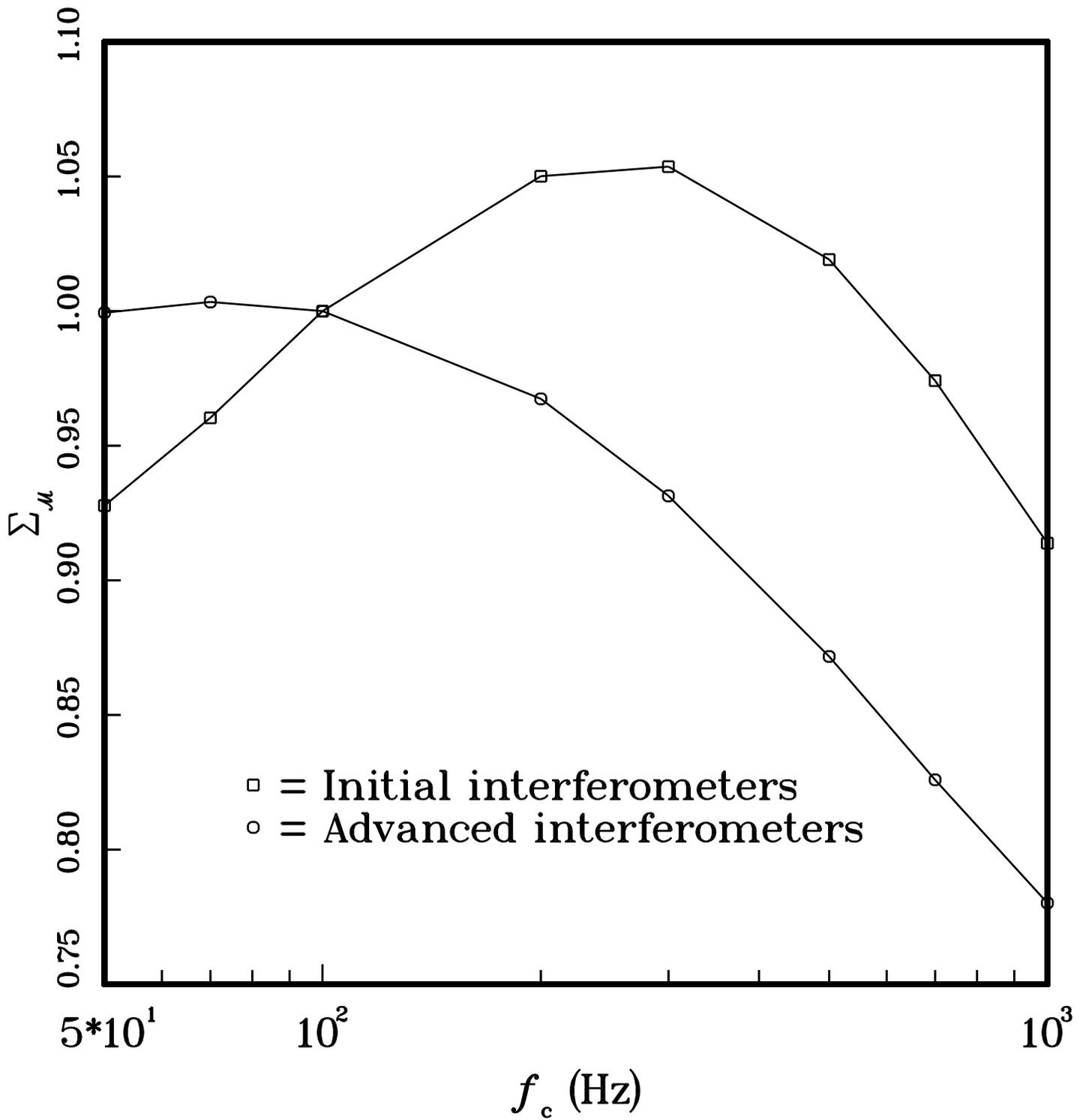
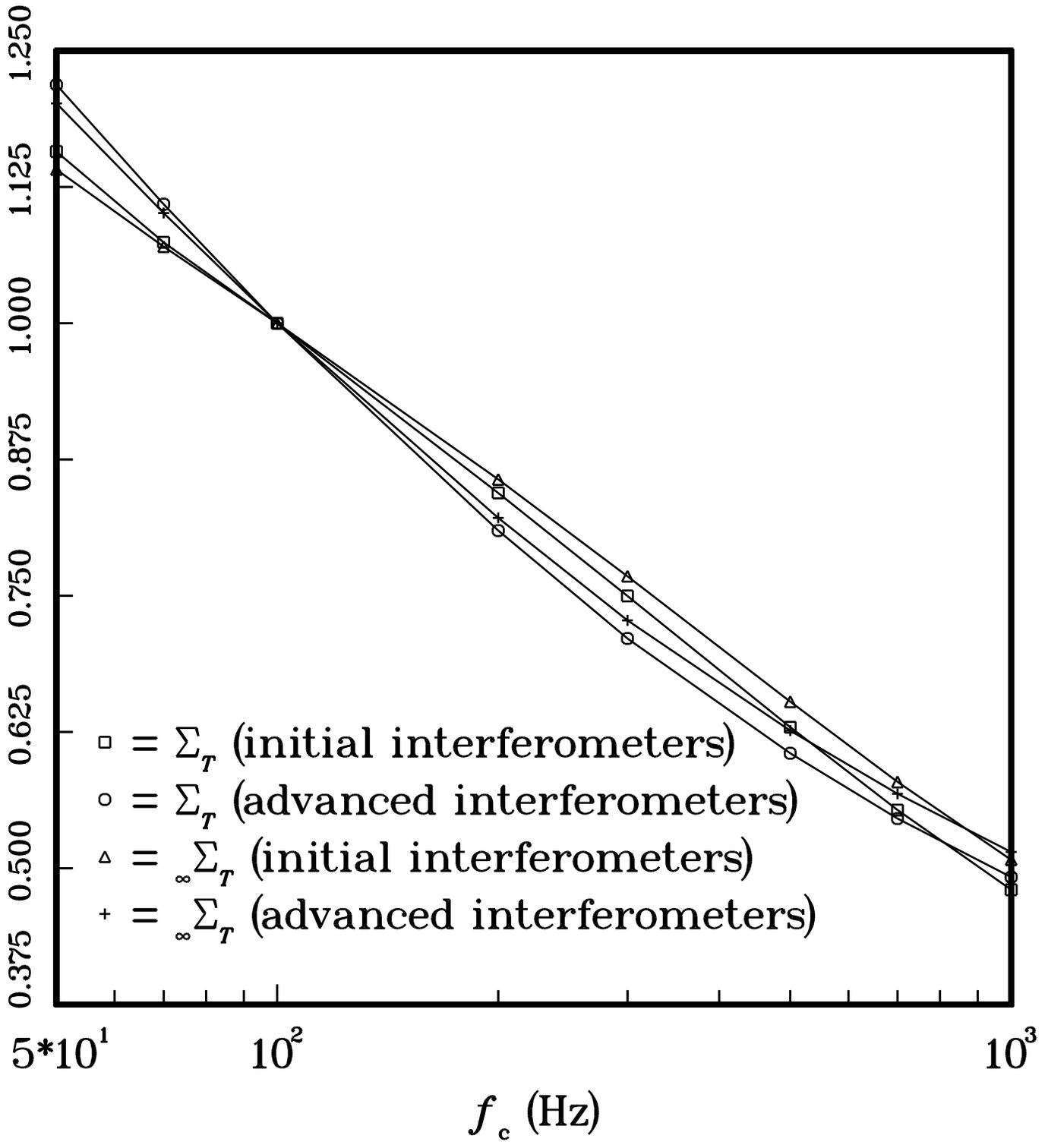
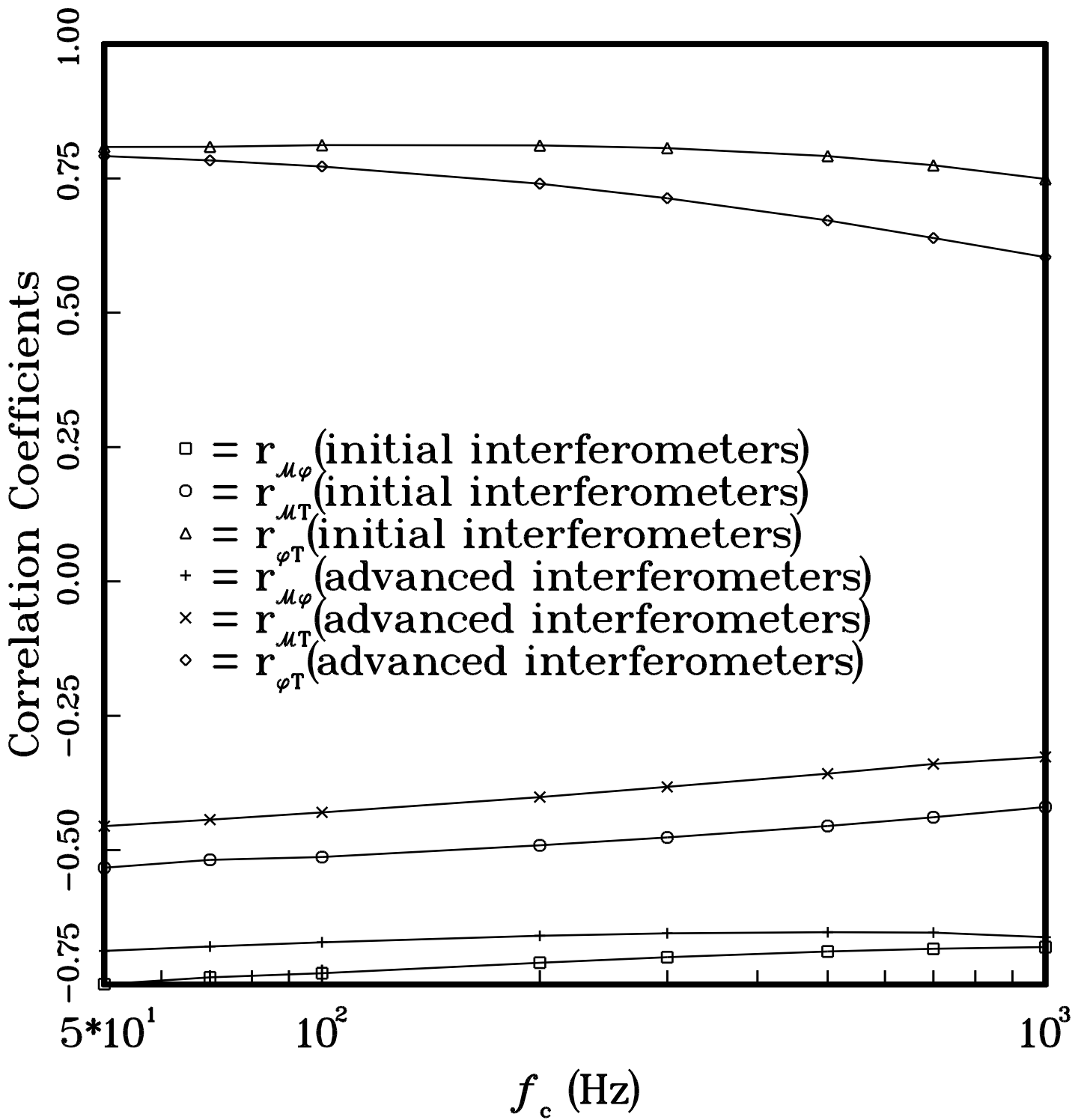
\end{document}